\documentclass[11pt,a4paper]{article}

\usepackage{jcappub}
\usepackage{graphicx}
\usepackage{amsmath}
\usepackage{amssymb}
\usepackage{subfigure}

\newcommand{\goodgap}{\hspace{\subfigtopskip} \hspace{\subfigbottomskip}}

\title{An updated analysis of two classes of $f(R)$ theories of gravity}

\author[1]{Vincenzo F. Cardone,}
\author[2,3,4,5]{Stefano Camera,}
\author[2,4,6]{and Antonaldo Diaferio}

\affiliation[1]{INAF, Osservatorio Astronomico di Roma, V. Frascati 33, 00040 Monte Porzio Catone (Roma), Italy}
\affiliation[2]{Dipartimento di Fisica Generale ``A. Avogadro'', Universit\`a degli Studi di Torino, V. Pietro Giuria 1, 10125 Torino, Italy}
\affiliation[3]{Dipartimento di Fisica Teorica, Universit\`a degli Studi di Torino, V. Pietro Giuria 1, 10125 Torino, Italy}
\affiliation[4]{INFN, Sezione di Torino, V. Pietro Giuria 1, 10125 Torino, Italy}
\affiliation[5]{CENTRA, Instituto Superior T\'ecnico, Universidade T\'ecnica de Lisboa, Av. Rovisco Pais 1, 1049-001 Lisboa, Portugal}
\affiliation[6]{Harvard-Smithsonian Center for Astrophysics, 60 Garden St., Cambridge, MA 02138, USA}

\emailAdd{{\tt winnyenodrac@gmail.com}}
\emailAdd{{\tt camera@ph.unito.it}}
\emailAdd{{\tt diaferio@ph.unito.it}}

\toccontinuoustrue

\abstract{The observed accelerated cosmic expansion can be a signature of fourth\,-\,order gravity theories, where the acceleration of the Universe is a consequence of departures from Einstein General Relativity, rather than the sign of the existence of a fluid with negative pressure. In the fourth\,-\,order gravity theories, the gravity Lagrangian is described by an analytic function $f(R)$ of the scalar curvature $R$ subject to the demanding conditions that no detectable deviations from standard GR is observed on the Solar System scale. Here we consider two classes of $f(R)$ theories able to pass Solar System tests and investigate their viability on cosmological scales. To this end, we fit the theories to a large dataset including the combined Hubble diagram of Type Ia Supernovae and Gamma Ray Bursts, the Hubble parameter $H(z)$ data from passively evolving red galaxies, Baryon Acoustic Oscillations extracted from the seventh data release of the Sloan Digital Sky Survey (SDSS) and the distance priors from the Wilkinson Microwave Anisotropy Probe seven years (WMAP7) data. We find that both classes of $f(R)$ fit very well this large dataset with the present\,-\,day values of the matter density, Hubble constant and deceleration parameter in agreement with previous estimates; however, the strong degeneracy among the $f(R)$ parameters prevents us from strongly constraining their values. We also derive the growth factor $g = d\ln{\delta}/d\ln{a}$, with $\delta = \delta \rho_M/\rho_M$ the matter density perturbation, and show that it can still be well approximated by $g(z) \propto \Omega_M(z)^{\gamma}$. We finally constrain $\gamma$ (on some representative scales) and investigate its redshift dependence to see whether future data can discriminate between these classes of $f(R)$ theories and standard dark energy models.}


\begin{document}

\maketitle

\section{Introduction}

To explain the present-time accelerated expansion of the Universe and the temperature anisotropy pattern of the Cosmic Microwave Background (CMB) radiation (e.g. \cite{Union2,WMAP7}), the current concordance cosmological model relies on the properties of two so-called ``dark components:'' Dark Matter (DM) and Dark Energy (DE). On the one hand, the need for the former -- a non-relativistic, weakly-interacting, non-baryonic matter -- matches today's requests for an extension of the standard model of particle physics: for instance, there are several proposals for DM candidates in the framework of supersymmetry \cite{JKS96,BHS04,F06}. However, different approaches have also been followed -- for example MoND (modified Newtonian dynamics) \cite{M83,SmG02,MondFP} and conformal gravity \cite{Mannheim1,Mannheim2}. On the other hand, DE still represents a difficult theoretical challenge (see \cite{Bianchi} for a different perspective). Indeed, there are strong fine-tunings problems, regardless of whether one interprets DE as a cosmological constant $\Lambda$, whose tiny but non-zero value is not supported by any geometrical symmetry, or whether one considers DE as the vacuum energy of a quantum field. In this case, the discrepancy between its actual measured value and the theoretical estimate is bigger than a hundred orders of magnitude.

To remove the DE problem, a different approach has recently started to be investigated. It is based on the consideration that $\Lambda$ was originally proposed as a constant term in the left-hand (geometric) side of Einstein's field equations. By generalising this approach, one can argue whether it is possible to reproduce the current cosmic accelerated expansion by adding a non-constant time-dependent term in Einstein's tensor. Actually, the effort of modifying and generalising the Hilbert-Einstein action of gravity dates back to just few years after Einstein's seminal papers (e.g. \cite{Schmidt:2006jt} for a historical review), and it has been also proposed by Starobinsky \cite{Starobinsky:1980te} in order to explain the cosmic inflation in the early Universe. This idea has again been suggested nowadays to correctly describe the current accelerated expansion of the Universe without any exotic fluid \cite{Capozziello:2002rd,Capozziello:2003gx,Nojiri:2003ft,Nojiri:2003wx,Carroll:2003wy,Capozziello:2005ku}. These modified-action theories of gravity are widely known as $f(R)$ theories, because in the gravity Lagrangian the Ricci scalar $R$ is replaced by the addition of a function $f(R)$, which can be, in principle, generic.

However, GR is a well-tested theoretical framework, at least with respect to the Solar System scale of distances. Therefore, any $f(R)$ theory which attempts to solve the late-time acceleration problem has to face the Solar System tests of gravity. Recently, two models carefully designed to pass the local gravity tests but still providing an accelerated cosmic expansion have been proposed \cite{Star07,HS07}. In this work, we analyse and test these viable $f(R)$ models. Specifically, we fit these theories against a large dataset which includes the combined Hubble diagram of Type Ia Supernovae and Gamma Ray Bursts, the Hubble parameter $H(z)$ data from passively evolving red galaxies, Baryon Acoustic Oscillations extracted from the seventh data release of the Sloan Digital Sky Survey (SDSS) and the distance priors from the seven years data of the Wilkinson Microwave Anisotropy Probe (WMAP7). Furthermore, we derive the growth factor $g = d\ln{\delta}/d\ln{a}$, with $\delta = \delta \rho_M/\rho_M$ the matter density perturbation.

%We find that both classes of $f(R)$ fit very well this large dataset with the present\,-\,day values of the matter density, Hubble constant and deceleration parameter in agreement with previous estimates. Unfortunately, the strong degeneracy among the $f(R)$ parameters prevents us from setting strong constraints on their values. We also outline how the growth factor can still be well approximated as $g(z) \propto \Omega_M(z)^{\gamma}$, and constrain $\gamma$ on some representative scales. To this aim, we investigate its redshift dependence to see whether future data can discriminate among these classes of $f(R)$ theories and standard dark energy models.

\section{$f(R)$ theories}

Being a straightforward generalization of Einstein GR, fourth order gravity theories have been investigated almost as soon as the original Einstein theory appeared. It is, therefore, not surprising that the corresponding field equations and the resulting cosmology have been so widely discussed in the literature. We here first review the basic of $f(R)$ theories and then introduce the two classes of $f(R)$ models we will investigate in this work.

\subsection{$f(R)$ cosmology}

In the framework of the metric approach, the field equations are obtained by varying the gravity action

\begin{equation}
S = \int{d^4x \sqrt{-g} \left [ \frac{f(R)}{2 \kappa} + {\cal{L}}_M \right ]}
\label{eq: fogaction}
\end{equation}
with respect to the metric components. We obtain

\begin{equation}
f^{\prime} R_{\mu \nu} - \nabla_{\mu \nu} f^{\prime} + \left ( \square f^{\prime} - \frac{1}{2} f \right ) g_{\mu \nu} = \kappa T_{\mu \nu} \ ,
\label{eq: fogfield}
\end{equation}
where $R$ is the scalar curvature, $\kappa = 8 \pi G$ (in units with $c=1$), ${\cal{L}}_M$ is the standard matter Lagrangian with $T_{\mu \nu}$ the matter stress\,-\,energy tensor, and the prime denotes derivative with respect to $R$. Note that, for $f(R) = R - 2 \Lambda$, one obtains the usual Einstein equations with a cosmological constant $\Lambda$. In the general case, a further scalar degree of freedom, conveniently represented by the scalar function $\phi = f^{\prime} - 1$, is introduced. The trace of Eq.(\ref{eq: fogfield}) gives the evolution for the effective field \cite{dev}

\begin{equation}
\square \phi = \frac{dV}{d\phi} + \frac{\kappa}{3} T
\label{eq: fieldtrace}
\end{equation}
with the potential $V$ related to $f(R)$ by

\begin{equation}
\frac{dV}{dR} = \frac{1}{3} (2 f - f^{\prime} R) f^{\prime \prime} \ .
\label{eq: fieldpot}
\end{equation}
In a spatially flat homogenous and isotropic universe, the time\,-\,time component of Eq.(\ref{eq: fogfield}) gives the evolution of the Hubble parameter, $H = \dot{a}/a$ (with the dot denoting derivative with respect to cosmic time $t$), being\,:

\begin{equation}
H^2 + \frac{d(\ln{f^{\prime})}}{dt} H + \frac{f - R f^{\prime}}{6 f^{\prime}} = \frac{\kappa \rho_M}{3 f^{\prime}} \ ,
\label{eq: hubblepar}
\end{equation}
while the trace equation (\ref{eq: fieldtrace}) may be recasted as a Klein\,-\,Gordon relation for $\phi$\,:

\begin{equation}
\ddot{\phi} + 3 H \dot{\phi} + \frac{dV}{d\phi} = \frac{\kappa}{3} \rho_M \ .
\label{eq: kglike}
\end{equation}
Eqs.(\ref{eq: hubblepar}) and (\ref{eq: kglike}) can be rearranged in such a way that a single equation for the Hubble parameter is obtained. To this end, it is first convenient to change variable from the time $t$ to the redshift $z = 1/a - 1$ and define the dimensionless curvature $\xi = R/m^2$, with

\begin{equation}
m^2 = \frac{\kappa^2 \rho_M(\eta = 0)}{3} \simeq (8315 \ {\rm Mpc})^{-2} \ \left ( \frac{\Omega_M h^2}{0.13} \right )
\label{eq: defm}
\end{equation}
a convenient curvature scale which depends on the present day values of the matter density parameter $\Omega_M$ and the Hubble constant $h = H_0/(100 \ {\rm km/s/Mpc})$. Assuming dust as gravity source and introducing $E = H(z)/H_0$, it is then only a matter of algebra to get\,:
\begin{equation}
E^2(z) = \frac{\Omega_M \left [ (1 + z)^3 + (\xi f^{\prime} - m^2 f)/6 \right ]} {f^{\prime} - m^2 (1 + z) (d\xi/dz)} \ ,
\label{eq: hubevsxi}
\end{equation}

\begin{equation}
m^2 f^{\prime \prime} \frac{d^2 \xi}{dz^2} + m^4 f^{\prime \prime \prime} \left ( \frac{d\xi}{dz} \right )^2 - \left [2 - \frac{d\ln{E}}{d\ln{(1 + z)}} \right ] \frac{m^2 f^{\prime \prime}}{1 + z} \frac{d\xi}{dz}
= \frac{\Omega_M}{E^2} \left [ (1 + z) - \frac{2 m^2 f - \xi f^{\prime}}{3 (1 + z)^2} \right ] \ .
\label{eq: xizeq}
\end{equation}
By inserting Eq.(\ref{eq: hubevsxi}) into Eq.(\ref{eq: xizeq}), we get a single second order nonlinear differential equation for $\xi(z)$ that can be solved numerically provided $f(R)$ and the initial conditions are given. To this end, we first remember that, because we are using the RW metric, it is

\begin{equation}
R = 6 (\dot{H} + 2 H^2)
\label{eq: defR}
\end{equation}
so that we get for the present-day values \cite{CCS08}\,:

\begin{displaymath}
R_0 = 6 H_0^2 (1 - q_0) \ \ , \ \ \dot{R}_0 = 6 H_0^3 (j_0 - q_0 - 2)  \ \ ,
\end{displaymath}
with $(q_0, j_0)$ the present-day values of the deceleration ($q = -H^{-2} \ddot{a}/a$) and jerk ($j = H^{-3} \dddot{a}/a$) parameters. It is then straightforward to get\,:

\begin{equation}
\left \{
\begin{array}{l}
\displaystyle{\xi(z = 0) = (6/\Omega_M) (1 - q_0)} \\
~ \\
\displaystyle{d\xi/dz(z = 0) = (6/\Omega_M) (j_0 - q_0 - 2)} \\
\end{array}
\right . \ .
\label{eq: incond}
\end{equation}
As a final remark, we note that, because of the definition of $\xi$, Eq.(\ref{eq: xizeq}) is a single fourth\,-\,order nonlinear differential equation for the scale factor $a(t)$ so that we need to know the values of the derivatives up to the third order to determine the evolution of $a(t)$. This explains why the jerk parameter $j_0$ is required.

\subsection{$f(R)$ models}

A key role in fourth\,-\,order gravity is clearly played by the functional expression of $f(R)$. Although such a choice is in principle arbitrary, there are some fundamental tests that have to be passed in order to get a physically viable theory. First, at the Solar System scales, $f(R)$ theories introduce a scalar degree of freedom which couples to matter and originates a long range fifth force that can lead to incorrect PPN parameters (see, e.g., \cite{Ch03,ESK06,CSE07,Ol07} and refs. therein). As a possible way out, one can invoke a chamaleon effect \cite{KW04a,KW04b} and tailor the $f(R)$ expression  in such a way to give rise to a mass squared term which is large and positive in high curvature environments \cite{Cem06,FTBM07,NvA07,CT08}. On the other hand, in the early Universe, one wants to recover the standard GR in order to preserve the agreement with the nucleosynthesis constraints. Among the possible choices left out by the above conditions, we will consider here two popular classes of $f(R)$ models which we briefly describe in the following.

After working out the above conditions for the viability of a $f(R)$ theory, Hu \& Sawicki first proposed the following functional expression \cite{HS07}\,:

\begin{equation}
f(R) = R - m^2 \frac{c_1 (R/m^2)^n}{1 + c_2 (R/m^2)^n}
\label{eq: frhs}
\end{equation}
where $m^2$ is given by (\ref{eq: defm}), and $(n, c_1, c_2)$ are positive dimensionless constants. Note that, since $f(R = 0) = 0$, there is formally no cosmological constant term. Nevertheless, since

\begin{displaymath}
\lim_{m^2/R \rightarrow 0}{f(R)} \simeq - \frac{c_1}{c_2} m^2 + \frac{c_1}{c_2^2} m^2 \left ( \frac{m^2}{R} \right )^n \ ,
\end{displaymath}
we still recover an effective $\Lambda$ term in high curvature $(m^2/R \rightarrow 0)$ environments. In particular, in the limit $R_0 >> m^2$ and $R_0 >> R_{\star}$, the expansion rate $H$, in the early universe, will be the same as in $\Lambda$CDM with an effective matter density parameter $\Omega_{M,eff} = 6 c_2/(c_1 + 6 c_2)$ that guarantees that the nucleosynthesis constraints are satisfied.

The HS model is determined by the three parameters $(n, c_1, c_2)$: two of them can be fixed in terms of observable quantities. To this end, we first note that evaluating Eq.(\ref{eq: hubevsxi}) at $z = 0$ gives a relation between the derivatives of $f$ and the present day value of $d\xi/dz$. Second, we expect that $f^{\prime}(R = R_0) = f_{R0}$ only mildly departs from the GR value $f_{R0} = 1$ in order to have an effective gravitational constant $G_{eff} = G/f^{\prime}$ as close as possible to the local one today. We therefore have\,:

\begin{equation}
\left \{
\begin{array}{l}
\displaystyle{\frac{\Omega_M [1 + (\xi_0 f_{R0} - m^2 f_0)/6]}{f_{R0} - m^2 f_0 \xi_{p0}} = 1} \\
~ \\
\displaystyle{f_{R0} = 1 - \varepsilon}
\end{array}
\right . \ ,
\label{eq: c1c2eq}
\end{equation}
with $\xi_0 = \xi(z = 0)$, $f_0 = f(R = R_0)$, $\xi_{p0} = d\xi/dz(z = 0)$, and $\varepsilon$ an additional free parameter. Inserting the corresponding expressions for the HS model, Eqs.(\ref{eq: c1c2eq}) can then be solved to express $(c_1, c_2)$ as a function of $(\Omega_M, q_0, j_0, \varepsilon)$ thus simplifying the choice of the parameters when fitting the model to the data.

Another possibility to satisfy all the constraints hinted at above is offered by the Starobinsky proposal \cite{Star07}\,:

\begin{equation}
f(R) = R + \lambda R_{\star} \left [ \left ( 1 + \frac{R^2}{R_{\star}^2} \right )^{-n} - 1 \right ]
\label{eq: frst}
\end{equation}
with $R_{\star}$ a scaling curvature parameter and $(\lambda, n)$ two positive constants. We will refer to this class of $f(R)$ theories as the St model. Note that, even in this case, $f(0) = 0$ so that no actual cosmological constant is present. Nevertheless, an effective one is recovered in the high curvature regime as can be seen from $f(R \gg R_{\star}) \sim R -2 \Lambda_{\infty}$ with $\Lambda_{\infty} = \lambda R_{\star}/2$. The St model parameters are $(n, R_{\star}, \lambda)$, but it is again convenient to reparametrize the model differently. We again resort to Eqs.(\ref{eq: c1c2eq}) inserting the corresponding expressions of $f$ and $f^{\prime}$ for the St model and solving them with respect to $(\lambda, R_{\star})$. Note that we thus use the same set of parameters to describe both the HS and the St models.

These two classes of $f(R)$ theories are actually quite similar at both very low and very high redshifts since they are both built up by imposing the same constraints on $f(R)$. Moreover, they both aim at mimicking the successful $\Lambda$CDM scenario in the late and early Universe. In other words, the HS and St models both reduces to the GR\,+\,$\Lambda$ case in the limits of very high and very low curvatures. What makes them different is the way the two extreme cases are connected, i.e. how the Universe evolves from the present day phase dominated by $\Lambda$ to the early matter epoch.

For completeness, we finally remind the reader that the two models we are considering here are not the only viable ones; other possible examples are given in \cite{AB07,SO07}. It is, moreover, possible to design $f(R)$ models that can also provide an inflationary expansion in the very early Universe \cite{SOinfl,ABS10}. However, all these other cases share many similarities with the HS and St models so that we are confident that exploring these two classes of fourth\,-\,order gravity theories should provide some general conclusions on their viability.

\subsection{The scale factor evolution}

To obtain the numerical solutions of Eqs.(\ref{eq: hubevsxi}) and (\ref{eq: xizeq}) more efficiently, we resort to an analytical approximation. We note that, for both the HS and St models, the gravity Lagrangian is excellently approximated by the GR\,+\,$\Lambda$ one by construction. Therefore, unless we are interested in the very early universe (where both $f(R)$ models have a vanishing $\Lambda$ term), we can expect that the dimensionless Hubble parameter, $E(z) = H(z)/H_0$, is close to $\Lambda$CDM
\begin{equation}
E(z >> z_{\Lambda}) \simeq E_{\lambda}(z) = \sqrt{\Omega_{M,eff} (1 + z)^3 + \Omega_{\gamma} (1 + z)^4 + (1 - \Omega_{M,eff} - \Omega_r)}
\label{eq: elambda}
\end{equation}
where $z_{\Lambda}$ is a characteristic redshift marking the transition to the $\Lambda$CDM regime, $\Omega_{\gamma}$ is the radiation density parameter and
\begin{equation}
\Omega_{M,eff} = \left \{
\begin{array}{ll}
\displaystyle{\frac{6 c_2}{c_1 + 6 c_2}} & {\rm \ \ for \ \ HS} \\
~ & ~ \\
\displaystyle{1 - \frac{\lambda R_{\star}}{6 H_0^2}} & {\rm \ \ for \ \ St} \\
\end{array}
\right .
\label{eq: omeff}
\end{equation}
is the effective matter density parameter as inferred from the effective $\Lambda$ term introduced in this limit by the HS and St models.

On the other hand, from the point of view of the background evolution, every modified gravity theory is equivalent to a dark energy (DE) model with a suitably reconstructed equation of state. Moreover, for a large class of DE theories, the equation of state is well approximated over a large redshift range by the phenomenological Chevallier\,-\,Polarski\,-\,Linder (CPL) \cite{CP01,L03} parametrization, $w(z) = w_0 + w_a (1 - a)$, leading to

\begin{equation}
E_{CPL}^2(z) = \Omega_M (1 + z)^3 + (1 - \Omega_M) (1 + z)^{3(1 + w_0 + w_a)} \exp{\left [ - \frac{3 w_a z}{1 + z} \right ]}
\label{eq: ecpl}
\end{equation}
for the dimensionless Hubble parameter so that, for $(w_0, w_a) = (-1, 0)$, $E_{\Lambda}(z)$ is recovered.

Motivated by these considerations, for both the HS and St models, we have therefore fitted the numerical solution of Eqs.(\ref{eq: hubevsxi}) and (\ref{eq: xizeq}) with the following ansatz\,:

\begin{equation}
E(z) = \left \{
\begin{array}{ll}
{\cal{E}}(z) E_{CPL}(z, \Omega_M) + [1 - {\cal{E}}(z)] E_{\Lambda}(z, \Omega_M) & z \le z_{\Lambda} \\
~ & ~ \\
E_{\Lambda}(z, \Omega_{M,eff}) & z \ge z_{\Lambda} \\
\end{array}
\right .
\label{eq: hubapprox}
\end{equation}
where
\begin{equation}
{\cal{E}}(z) = \sum_{i = 1}^{3}{e_i (z - z_{\Lambda})^{i}}
\label{eq: defepsfun}
\end{equation}
is an interpolating function with $e_i$ and $z_\Lambda$ fitting parameters. This approximating function excellently reproduces the numerical solution whatever the model parameters $(\Omega_M, q_0, j_0, n, \varepsilon)$ are for both the HS and St models with a rms error which is far lower than $0.1\%$ over the full redshift range $(0, 1000)$. A simple look at the $f(R)$ functions makes quite easy to understand why this happens. Indeed, both $f(R)$ functions converges to $f(R) = R - 2 \Lambda_{eff}$ as $z$ increases so that it is not surprising that, after a critical value $z_{\Lambda}$ when $f(R)$ is indistinguishable from the $\Lambda$CDM Lagrangian, the Hubble parameter becomes exactly the same with the value of the effective cosmological constant depending on the model parameters. On the other hand, at very low redshift, the CPL parametrization does an excellent job in mimicking $E(z << 1)$ as expected. The interpolating function ${\cal{E}}(z)$ then only smooths the transition between the two regimes in an efficient way.

A subtle remark is in order here concerning the value of $\Omega_M$. Indeed, while for $z \le z_{\Lambda}$, we use the actual matter density parameter (and neglect radiation for simplicity), the effective one enters the $z \ge z_{\Lambda}$ approximation. Therefore, a discontinuity in $z_{\Lambda}$ is formally present in our approximation. Actually, it is easy to show that, for all reasonable model parameters, $\Omega_M$ and $\Omega_{M,eff}$ are almost perfectly equal so that the discontinuity can not be detected at all and $E(z)$ is, to all extents, a continuous function. On the other hand, it is worth stressing that $(w_0, w_a, e_1, e_2, e_3, z_{\Lambda})$ depend in a complicated way on the $f(R)$ model parameters. Actually, by numerical attempts, we discovered that, within a very good approximation, $(e_1, e_2, e_3)$ are the same for all the models considered so that the four remaining quantities $(q_0, j_0, n, \varepsilon)$ collapse into three parameters only, namely $(w_0, w_a, z_{\Lambda})$. We can therefore anticipate that strong degeneracies among them will take place since different sets will give rise to the same $E(z)$. Moreover, $z_{\Lambda}$ increases with $\varepsilon$ because the smaller is $\varepsilon$, the closer is $f_{R0}$ (and hence $f^{\prime}$) to 1, i.e. the smaller $\varepsilon$ is the quicker is the convergence to the $\Lambda$CDM Lagrangian. As a consequence, the smaller is $\varepsilon$, the narrower will be the redshift range where $E(z)$ departs from the concordance model.

\section{Constraining $f(R)$ models}

The HS and ST models provide an accelerated expansion in a matter only universe because, by construction, they mimic the successful $\Lambda$CDM scenario in both the low and high redshift regimes. Nevertheless, fitting the model to the data is still of significant help to constrain their wide parameter space and pin down their predictions on other not fitted quantities. In order to lift the degeneracies among the model parameters, one can not rely on low redshift probes only, but has to add further data tracing higher $z$ or based on quantities related in a different way to the Hubble parameter. Such considerations motivated our choice of the dataset described below.

\subsection{Observational data}

We use the 557 SNeIa in the Union2 \cite{Union2} sample to probe the evolution of the low redshift universe (over the range $0.015 \le z \le 1.37$). Neglecting systematics, the SNeIa likelihood term will simply read
\begin{equation}
{\cal{L}}_{SNeIa}({\bf p}) \propto \exp{[-\chi_{SNeIa}^2({\bf p})/2]}
\label{eq: likesneia}
\end{equation}
with
\begin{equation}
\chi_{SNeIa}^2({\bf p}) = \sum_{i = 1}^{{\cal{N}}_{SNeIa}}{\left [ \frac{\mu_{obs}(z_i) - \mu_{th}(z_i, {\bf p})}{\sigma_i} \right ]^2}
\label{eq: defchisneia}
\end{equation}
where $\sigma_i$ is the error on the observed distance modulus $\mu_{obs}(z_i)$ for the i\,-\,th object at redshift $z_i$ and the theoretical distance modulus is given by
\begin{equation}
\mu_{th}(z, {\bf p}) = 25 + 5 \log{\left [ \frac{c}{H_0} (1 + z) r(z, {\bf p}) \right ]}
\label{eq: defmuth}
\end{equation}
with $r(z)$ the dimensionless comoving distance
\begin{equation}
r(z, {\bf p}) = \int_{0}^{z}{\frac{dz'}{E(z', {\bf p})}} \ .
\label{eq: defrz}
\end{equation}
and ${\bf p}$ the set of model parameters for the assumed $f(R)$ class of theories.

The HS and St models differ from each other in the intermediate redshift regime and we need to trace the Hubble diagram in the matter dominated era to discriminate between them. Thanks to the enormous energy release that makes them visibile up to $z \sim 8.2$, Gamma Ray Bursts (GRBs) stand out as good candidates to this task \cite{Schaefer,CCD09}. We therefore use the GRBs Hubble diagram as recently derived in \cite{Marcy} (see also \cite{WQD11}) for the catalog of 115 GRBs in \cite{XS09}. In order to use a model independent calibration of the GRBs scaling relations (but see \cite{DOC11} for the limits of these assumption), we use the results in \cite{Marcy} obtained using the local regression method (see \cite{CCD09} and refs. therein for details). From the sample, we remove the objects with $z \le 1.4$ in order to avoid correlations, difficult to quantify, with the SNeIa data introduced by the calibration method. As a result, we end up with 64 objects with $1.48 \le z \le 5.60$, so probing the Hubble diagram deep into the matter dominated era.

Note that, when using GRBs, the likelihood term will be the same as Eq.(\ref{eq: likesneia}), but with the denominator in Eq.(\ref{eq: defchisneia}) changed to $(\sigma_i^2 + \sigma_{int}^2)^{1/2}$ with $\sigma_{int}$ the unknown intrinsic scatter of the GRBs around the theoretical Hubble diagram. This term may come out from different sources. First, the empirical GRBs Hubble diagram has been obtained by averaging the distance modulus of each object as inferred from multiple scaling relations so that, if an object is an outlier for one of the correlations used, its distance modulus will be shifted from the true one. Moreover, the calibration procedure may also introduce some bias due to neglecting any redshift evolution of the scaling relations coefficients which can not be excluded given the wide redshift range probed. Therefore, we leave $\sigma_{int}$ as an unknown parameter\footnote{Note that a similar term is also present for SNeIa, but it is estimated to be $\sigma_{int} = 0.15$ and yet included into the error $\sigma_i$ provided in the Union2 dataset.} and marginalize over it in the likelihood analysis.

While both SNeIa and GRBs are based on the concept of standard candles, an alternative way to probe the background evolution of the universe relies on the the use of standard rulers. A nowadays widely used example is represented by the Baryonic Acoustic Oscillations (hereafter, BAOs) which are related to the imprint of the primordial acoustic waves on the galaxy power spectrum. In order to use BAOs as constraints, we follow \cite{P10} by first defining\,:
\begin{equation}
d_z = \frac{r_s(z_d)}{D_V(z)}
\label{eq: defdz}
\end{equation}
with $z_d$ the drag redshift (computed using the approximated formula in \cite{EH98}), $r_s(z)$ the comoving sound horizon given by\,:
\begin{equation}
r_s(z) = \frac{c}{\sqrt{3}} \int_{0}^{(1 + z)^{-1}}{\frac{da}{a^2 H(a) \sqrt{1 + (3/4) \Omega_b/\Omega_{\gamma}}}} \ ,
\label{eq: defsoundhor}
\end{equation}
and $D_V(z)$ the volume distance defined by \cite{Eis05}\,:
\begin{equation}
D_V(z) = \left \{ \frac{c z}{H(z)} \left [ \frac{D_L(z)}{1 + z} \right ]^2 \right \}^{1/3} \ .
\label{eq: defdv}
\end{equation}
We finally constrain the model parameters by introducing the likelihood function
\begin{equation}
{\cal{L}}_{BAO}({\bf p}) = \frac{\exp{\left [ - ({\bf D}_{BAO} \cdot {\bf C}_{BAO}^{-1} \cdot {\bf D}_{BAO}^T)/2 \right ]}}
{2 \pi |{\bf C}_{BAO}^{-1}|^{1/2}} \ ,
\label{eq: deflikebao}
\end{equation}
with ${\bf D}_{BAO}^T = (d_{0.2}^{obs} - d_{0.2}^{th}, d_{0.35}^{obs} - d_{0.35}^{th})$ and ${\bf C}_{BAO}$ the BAO covariance matrix. The values of $d_z$ at $z = 0.20$ and $z = 0.35$ have been estimated by \cite{P10}; we also use the values of the observed $d_z$ and ${\bf C}_{BAO}$ provided by \cite{P10}.

Both the Hubble diagram and the BAO are distance related quantities so that they probe the integrated expansion rate. Thus, the details of $H(z)$ are partially smoothed out so that severe degeneracies arise. In order to alleviate this problem, one should rely on direct estimates of $H(z)$ which can be obtained by noting that, from the relation $dt/dz = -(1+z) H(z)$, a measurement of $dt/dz$ at different $z$ gives, in principle, the Hubble parameter. This differential age method \cite{JL02} works best using fair samples of passively evolving galaxies with similar metallicity and low star formation rate so that they can be taken to be the oldest objects at a given $z$. Stern et al. \cite{S10II} have then used red envelope galaxies as cosmic chronometers determining their ages from high quality Keck spectra and applied the differential age method to estimate $H(z)$ over the redshift range $0.10 \le z \le 1.75$ \cite{S10I}. We use their data as input to the following likelihood function\,:

\begin{equation}
{\cal{L}}_H({\bf p}) = \frac{\exp{\left [ - \chi_H^2({\bf p})/2 \right ]}}{(2 \pi)^{{\cal{N}}_H/2} |{\bf C}_{H}^{-1}|^{1/2}} \ ,
\label{eq: deflikehz}
\end{equation}
with

\begin{equation}
\chi^2_H = \sum_{i = 1}^{{\cal{N}}_H}{\left [ \frac{H_{obs}(z_i) - H(z_i, {\bf p})}{\sigma_i} \right ]^2}
\label{eq: defchihz}
\end{equation}
where the sum is over the ${\cal{N}}_H = 11$ sample points and ${\bf C}_H$ is the diagonal covariance matrix.

The set of data described so far probes the evolution of the universe only during the late (dark energy dominated) era and the matter epoch, but it tells us nothing about the early Universe. This problem can be healed by adding the CMBR data to have a picture of the Universe at recombination $(z \simeq 1100)$. Komatsu et al. \cite{WMAP5} have shown that most of the information in the WMAP power spectrum may be summarized in the so called {\it distance priors}, i.e. constraints on\,: $(1)$ the redshift $z_{LS}$ of the last scattering surface (computed with the approximated formula in \cite{HS96}), $(2)$ the acoustic scale \cite{BET97,EB99,Page03}\,:
\begin{equation}
l_A = \frac{\pi (c/H_0) r(z_{LS})}{r_s(z_{LS})} \ ,
\label{eq: defla}
\end{equation}
and $(3)$ the shift parameter \cite{BET97,EB99,Page03}\,:
\begin{equation}
{\cal{R}} = \sqrt{\Omega_M} r(z_{LS}) \ .
\label{eq: defshift}
\end{equation}
We can then define the likelihood fucntion
\begin{equation}
{\cal{L}}_{CMB}({\bf p}) = \frac{\exp{\left [ - ({\bf D}_{CMB} \cdot {\bf C}_{CMB}^{-1} \cdot {\bf D}_{CMB}^T)/2 \right ]}}
{(2 \pi)^2 |{\bf C}_{CMB}^{-1}|^{1/2}} \ ,
\label{eq: deflikecmbr}
\end{equation}
with ${\bf D}_{CMB}$ the vector with the values of the differences between the observed and the theoretically predicted distance priors and ${\bf C}_{CMB}$ the corresponding covariance matrix. We rely on the seventh data release of the Wilkinson Microwave Anisotropy Probe (WMAP7) \cite{WMAP7} to set the observed distance priors and their covariance matrix.

\subsection{Likelihood analysis}

In the Bayesian approach to model testing, the parameter space of a given model is constrained by examining the region where a user\,-\,defined likelihood function ${\cal{L}}({\bf p})$ takes non negligible values. In particular, the best fit parameters are the set ${\bf p}_{bf}$ maximimizing ${\cal{L}}({\bf p})$, while the constraints on the i\,-\,th quantity $p_i$ are obtained by marginalizing ${\cal{L}}$ over all the parameters but $p_i$ itself. Mathematically, one defines\,:

\begin{equation}
{\cal{L}}_i(p_i) = \int{dp_1 \ldots \int{dp_{i - 1}} \int{dp_{i + 1} \ldots \int{dp_n {\cal{L}}({\bf p})}}}
\label{eq: defmarglike}
\end{equation}
and estimates the median value and the $68$ and $95\%$ confidence ranges for $p_i$ from ${\cal{L}}_i(p_i)$. As a general remark, we warn the reader that, in the context of Bayesian statistics, the best fit model represents the most plausible model in an Occam's razor sense given the data at hand. However, in a Bayesian context, the best fit parameters individually do not necessarily have to be probable, but rather they must have a high joint probability density that might occupy only a small volume of the parameter space. This situation can arise if the best fit solution does not lie in the bulk of the posterior probability distribution. Such a case may often occur when the posterior is non-symmetric in a high dimensional space so that the volume can dramatically increase with the distance from the best fit solution. In this case, the best fit solution for each parameter could easily lie outside the bulk of the individual posterior distribution for $p_i$ obtained by marginalizing over the other parameters. This is indeed what happens for our models so that we have preferred to remind the reader that this somewhat counterintuitive outcome is not a problem, but rather a common feature in statistics in multidimensional spaces.

The likelihood ${\cal{L}}({\bf p})$ takes the available data and the prior information on the model into account. In order not to bias our analysis, we include only flat priors on $(n, \log{\varepsilon})$ as explained later. We can therefore set\,:

\begin{equation}
{\cal{L}}({\bf p}) = \frac{\exp{\left [ - (h_{S} - h)^2/2 \sigma_{S}^2 \right ]}}{\sqrt{2 \pi \sigma_{S}^2}} \times
\Pi_{j}{{\cal{L}}_{(j)}({\bf p})}
\label{eq: deflike}
\end{equation}
where the first factor accounts for the recent determination of the Hubble constant from local distance calibrators by the SHOES collaboration which give $(h_S, \sigma_S) = (0.742, 0.036)$ \cite{shoes}, while $(j)$ stands for ${SNeIa, GRB, H, BAO, CMB}$ according to the definitions given in the previous subsections.

As a preliminary task, one should ask what the number of parameters to constrain is. First, we note that, in order to solve Eqs.(\ref{eq: hubevsxi}) and (\ref{eq: xizeq}), we must know the values of five quantities, namely $(\Omega_M, q_0, j_0, n, \log{\varepsilon})$, where we have moved to a logarithmic value for $\varepsilon$ since $\varepsilon$ is expected to be very small. A sixth parameter is the Hubble constant $h$ which enters through both the prior in Eq.(\ref{eq: deflike}) and the analysis of the $H(z)$ data. When including GRBs in the total likelihood, we need to estimate also the intrinsic scatter $\sigma_{int}$ in order to get the errors on the individual GRBs distance moduli. A further quantity is the physical baryon density $\omega_b = \Omega_b h^2$ entering directly through the WMAP7 distance priors and indirectly through its effect on the determination of $(z_d, z_{LS})$. Finally, the radiation term can not be neglected in the early universe so that we add this contribution when using the distance priors as constraints. Summarizing, the full set of parameters we consider reads\,:

\begin{displaymath}
{\bf p} = (\Omega_M, \Omega_{\gamma}, h, q_0, j_0, n, \log{\varepsilon}, \sigma_{int})
\end{displaymath}
while we set $\omega_b = 2.267 \times 10^{-2}$ \cite{WMAP5}. We use flat priors on all the parameters choosing large enough ranges to avoid biasing the likelihood analysis by any theoretical prejudice. We have, however, make two exceptions to this rule forcing $n \ge 1$ so that the chamaleon effect can take place \cite{CT08} and only considering models with\,:

\begin{displaymath}
-13 \le \log{\varepsilon} \le -3.0
\end{displaymath}
where the lower limit is simply set because of loss of sensitivity for smaller values, while the upper one is obtained by considering that, in order not to have significant deviations from the Newtonian gravitational potential on galactic scales, the constraint \cite{HS07}

\begin{equation}
\varepsilon \le 2 \times 10^{-6} \left ( \frac{v_{max}}{300 \ {\rm km/s}} \right ) \ ,
\label{eq: frgalaxy}
\end{equation}
with $v_{max}$ the maximum rotation velocity of the stars, must be satisfied. This would give us $\log{\varepsilon} \le -6$, but to be conservative we will extend the above range to include higher velocity systems (such as clusters of galaxies). Moreover, since $z_{LS}$ is actually not very sensitive to the model parameters (so that $z_{LS} \simeq 1090$ could be used for all the models within an excellent approximation), we do not expect to put strong constraints on $\Omega_{\gamma}$ and we can marginalize over it. Finally, we also marginalize over $\sigma_{int}$ since this nuisance quantity has not a straightforward physical interpretation being most related to the calibration of GRBs scaling relations rather to any physical effect.

In order to explore the eight dimensional parameter space, we use a Monte Carlo Markov Chain (MCMC) algorithm running three parallel chains of $\sim 25000$ points each and checking convergence with the usual Gelman\,-\,Rubin statistics \cite{Gelman1992}. After cutting out the burn in phase and thinning to avoid spurious correlations, the final merged chain also allows us to evaluate the constraints on some derived quantities of interest. To this end, we evaluate the parameter density distribution $\Theta({\bf p})$ along the chain and study the histogram of the values thus obtained to infer the corresponding median and confidence ranges.

\subsection{Results}

Since we have used the same parameters for defining both the HS and St models, it is worth discussing the constraints on these quantities in parallel thus also checking if they depend on the assumed $f(R)$ functional expression. In particular, the best fit values turn out to be\,:

\begin{displaymath}
(\Omega_M, h, q_0, j_0, n, \log{\varepsilon}) = (0.276, 0.719, -0.585, 0.995, 1.53, -5.95) \ ,
\end{displaymath}
for the HS model, and

\begin{displaymath}
(\Omega_M, h, q_0, j_0, n, \log{\varepsilon}) = (0.273, 0.725, -0.599, -0.058, 1.34, -10.02) \ ,
\end{displaymath}
for the St one. How well the best fit models reproduce the data may be quantitatively judged noting that, for the HS model, it is (with $\tilde{\chi}^2 = \chi^2/d.o.f$)\,:

\begin{displaymath}
\tilde{\chi}^2_{SNeIa} = 0.99 \ \ , \ \ \tilde{\chi}^2_{GRB} = 1.19 \ \ , \ \ \tilde{\chi}^2_{H} = 2.58 \ \ ,
\end{displaymath}
\begin{displaymath}
(d_{0.2}, d_{0.35}) = (0.1893, 0.1137) \ \ , \ \ (\ell_A, {\cal{R}}, z_{LS}) = (302.29, 1.725, 1091.5) \ \ ,
\end{displaymath}
while the St best fit model gives\,:

\begin{displaymath}
\tilde{\chi}^2_{SNeIa} = 0.99 \ \ , \ \ \tilde{\chi}^2_{GRB} = 1.21 \ \ , \ \ \tilde{\chi}^2_{H} = 2.56 \ \ ,
\end{displaymath}
\begin{displaymath}
(d_{0.2}, d_{0.35}) = (0.1901, 0.1142) \ \ , \ \ (\ell_A, {\cal{R}}, z_{LS}) = (302.36, 1.723, 1091.5) \ \ .
\end{displaymath}
Comparing with the observed values, we can therefore safely conclude that the both the HS and St models are in good agreement with the data. A cautionary note is in order here concerning the $\chi^2$ values reported above and hereafter. The MCMC code maximizes the likelihood (\ref{eq: deflike}) which is strictly not the same as minimizing the single reduced $\chi^2$ entering its definition.  Moreover, from a statistical point of view, the significance level of the above reduced $\chi^2$ values can not be estimated from the usual tables since these standard results do not take into account systematic errors or intrinsic scatter. It is also worth stressing that we have defined above the reduced $\chi^2$ computing the number of degrees of freedom as ${\cal{N}}_d - {\cal{N}}_p$ with ${\cal{N}}_d$ (${\cal{N}}_p$) the number of datapoints (parameters). While this is formally correct, some caution is needed when high values are obtained. Such large numbers may indicate that some of the parameters are actually unnecessary to fit a particular subset of the data. For instance, $\sigma_{int}$ does not enter at all in the fit to the SNeIa Hubble diagram and the $H(z)$ measurement, but it is nevertheless included in ${\cal{N}}_p$ when normalizing the $\chi^2$ for these datasets\footnote{Note that, although we finally marginalize over both $\Omega_{\gamma}$ and $\sigma_{int}$, they are nevertheless varied in the fitting procedure so that they must be included in the ${\cal{N}}_p$ value used to normalize the $\chi^2$.}. In particular, the high $\tilde{\chi}^2_H$ value is simply due to the inclusion of both $\sigma_{int}$ and $\Omega_{\gamma}$ which do not affect at all the fit to this subset. Indeed, the best fit models closely follows the $H(z)$ data even if $\tilde{\chi}^2_H$ is quite large. As a general remark, we therefore warn the reader to not overrate the reduced $\chi^2$ values.

\begin{table}[t]
\begin{center}
\begin{tabular}{ccccccc}
\hline
$x$ & \multicolumn{3}{c}{HS model}  & \multicolumn{3}{c}{St model} \\
\hline
~ & $x_{BF}$ & $\langle x \rangle$ & $(x_{med})_{-1 \sigma \ -2 \sigma}^{+1 \sigma \ +2 \sigma}$ & $x_{BF}$ & $\langle x \rangle$ & $(x_{med})_{-1 \sigma \ -2 \sigma}^{+1 \sigma \ +2 \sigma}$ \\
\hline \hline
~ & ~ & ~ & ~ & ~ & ~ & ~ \\
$\Omega_M$ & 0.276 & 0.279 & $0.279_{-0.014 \ -0.026}^{+0.014 \ +0.026}$ & 0.273 & 0.275 & $0.275_{-0.011 \ -0.024}^{+0.018 \ +0.029}$ \\
~ & ~ & ~ & ~ & ~ & ~ & ~ \\
$h$ & 0.719 & 0.720 & $0.720_{-0.017 \ -0.032}^{+0.017 \ +0.034}$ & 0.725 & 0.724 & $0.724_{-0.017 \ -0.032}^{+0.017 \ +0.032}$ \\
~ & ~ & ~ & ~ & ~ & ~ & ~ \\
$q_0$ & -0.585 & -0.585 & $-0.586_{-0.023 \ -0.043}^{+0.024 \ +0.043}$ & -0.599 & -0.655 & $-0.608_{-0.152 \ -0.491}^{+0.032 \ +0.054}$ \\
~ & ~ & ~ & ~ & ~ & ~ & ~ \\
$j_0$ & 0.995 & 1.036 & $1.020_{-0.022 \ -0.069}^{+0.053 \ +0.222}$ & -0.058 & 1.031 & $1.141_{-4.278 \ -8.936}^{+2.757 \ +13.185}$ \\
~ & ~ & ~ & ~ & ~ & ~ & ~ \\
$n$ & 1.53 & 1.67 & $1.41_{-0.31 \ -0.39}^{+1.12 \ +1.70}$ & 1.34 & 2.68 & $2.22_{-0.77 \ -1.06}^{+2.11 \ +2.68}$ \\
~ & ~ & ~ & ~ & ~ & ~ & ~ \\
$\log{\varepsilon}$ & -5.95 & -5.26 & $-4.66_{-2.76 \ -5.95}^{+1.18 \ +1.68}$ & -10.03 & -7.57 & $-8.59_{-1.17 \ -1.81}^{+4.09 \ +4.91}$ \\
~ & ~ & ~ & ~ & ~ & ~ & ~ \\
\hline
\end{tabular}
\end{center}
\caption{Constraints on the HS and St model parameters. For each parameter, we give the best fit, mean and median values and the $68$ and $95\%$ confidence ranges.}
\label{tab: fitconstr}
\end{table}

The comparison of the best fit models shows that both models predict the same matter abundance and quite similar values for the present day Hubble constant and deceleration parameter, while strikingly different results are obtained for the jerk parameter with the HS model giving almost the same value as predicted for the concordance $\Lambda$CDM scenario (i.e., $j_0 = 1$). However, one should better rely on the median value for the $j_0$ parameter since this latter takes fully into account the shape of the $j_0$ distribution. As Table\,\ref{tab: fitconstr} shows, the median $j_0$ values for the HS and St models are fully with the $68\%$ confidence ranges well overlapped. Moreover, the constraints in Table\,\ref{tab: fitconstr} are in good agreement with those in the literature (see, e.g., \cite{dev,matteo,ali,matteoter} and refs. therein). To this regard, it is worth stressing that our analysis differs from the previous ones in two important aspects. First, we use a more recent compilation of data (both SNeIa and BAO observational constraints) and add the GRB Hubble diagram to probe the matter dominated epoch. Second, we leave all the model parameters free to vary thus exploring the parameter space without imposing any limitation a priori such as, e.g., fixing $\Omega_M$ (as in \cite{matteo,matteoter}). Needless to say, such an approach does not come for free, the cost to pay being the rise of severe degeneracies in the 6D parameter space. An example is, indeed, provided by the $j_0$ distribution for the St model with a best fit value strongly different from the median one. However, the $\Lambda$CDM value $j_0 = 1$ is well within the $68\%$ confidence ranges for both the HS and St models so that, from the point of view of the cosmographic parameters, the constraints are still too weak to discriminate between the two models. Unexpectedly, the constraints on $(q_0, j_0)$ for the St model are much weaker than for the HS case. Quite weak constraints are obtained in both cases for $n$ and $\log{\varepsilon}$. In particular, the confidence ranges for $n$ are strongly asymmetric due to the hard prior $n > 1$ we have set to allow for a chamaleon\,-\,like effect to take place. In order to explore why this effect takes place, we can first consider the constraints on the original quantities entering the two $f(R)$ expressions. For the HS model, we find (median value and $68$ and $95\%$ confidence range)\,:

\begin{displaymath}
\log{c_1} = 3.47_{-0.91 \ -1.92}^{+2.01 \ +3.26} \ \ , \ \
\log{c_2} = 2.28_{-0.90 \ -1.91}^{+2.01 \ +3.30} \ \ ,
\end{displaymath}
so that Eq.(\ref{eq: frhs}) effectively reduces to $f(R) \simeq R - m^2 c_1/c_2$. The HS model is therefore quite similar to $\Lambda$CDM; this explains why the constraints on $j_0$ give values narrowly centred around the $\Lambda$CDM $j_0$. For the St model, we get\,:

\begin{displaymath}
\log{\lambda} = 1.50_{-0.93 \ -1.28}^{+1.31 \ +2.27} \ \ , \ \
\log{(R_{\star}/R_0)} = -1.73_{-1.27 \ -2.24}^{+0.93 \ +2.21} \ \ .
\end{displaymath}
so that, in the regime $R \simeq R_0$, the effective Lagrangian reduces to $f(R \simeq R_0) \simeq R_0 + \lambda R_0 \kappa (1 - \kappa^{2n})$ with $\kappa = R_{\star}/R_0$. Since $\kappa << 1$, we still have an effective cosmological constant term; however, for large values of $n$ and not negligible values of $\kappa$, the second term in parentheses, $\kappa^{2n}$, might not be negligible. Therefore, values of $j_0 \ne 1$ are possible and this explains the large confidence ranges.

Finally, as an alternative way to describe the background evolution for the HS and St models, we can work out the effective DE equation of state as\,:

\begin{equation}
1 + w_{eff}(z) = \left [ \frac{2}{3} \frac{d\ln{E(z)}}{d\ln{(1 + z)}} -
\frac{\Omega_M (1 + z)^3}{E^{2}(z)} \right ] \nonumber \\
\left [ 1 - \frac{\Omega_M (1 + z)^3}{E^{2}(z)} \right ]^{-1} \ .
\label{eq: eoseff}
\end{equation}
The effective EoS for both $f(R)$ models may be anticipated considering that the dimensionless Hubble parameter $E(z)$ may be well approximated by Eq.(\ref{eq: hubapprox}). Whichever are the model parameters, the $f(R)$ function reduces to $f_{\Lambda}$ at high $z$ so that the EoS reduces to $w_{eff}(z >> 1) = -1$ identically. On the other hand, at low $z$, both HS and St models approximate well the $\Lambda$CDM evolutionary history so that we expect only small deviations. That this is indeed the case is confirmed by the values of $w_0 = w_{eff}(z = 0)$ and $w_1 = dw_{eff}/dz(z = 0)$. Evaluating these quantities along the chain, we get for the HS model\,:

\begin{displaymath}
(1 + w_0) \times 10^{-3} = -1.9_{-5.5 \ -19.7}^{+2.1 \ +2.2} \ \ , \ \
w_1 \times 10^{-2} = 1.2_{-1.2 \ -6.2}^{+3.6 \ +13.9} \ \ ,
\end{displaymath}
so that we actually find only tiny deviations from the $\Lambda$CDM values $(w_0, w_1) = (-1, 0)$. On the contrary, more freedom is allowed by the St model. In this case, we obtain\,:

\begin{displaymath}
(1 + w_0) \times 10^{-1} = -0.01_{-1.6 \ -4.8}^{+0.02 \ +0.04} \ \ , \ \
w_1 = -0.12_{-4.1 \ -8.8}^{+3.0 \ +10.7} \ \ ,
\end{displaymath}
As it is apparent, an effective cosmological constant is still well consistent with these constraints, but also not negligible deviations from $w_0 = -1$ and a varying EoS are allowed. Since the jerk parameter depends on both $w_0$ and $w_1$ (see, e.g., \cite{CCS08} for its expression for the CPL model), we expect to be able to set only weak constraints on $j_0$ when fitting the St model to the data, while the opposite conclusion may be drawn for the HS case. % considering the above value of $w_1$ consistent with what we indeed get for $j_0$ confidence ranges in Table\,\ref{tab: fitconstr}.

\section{Time related observable quantities}

To constrain the cosmological parameters,
%Apart for the $H(z)$ data, all the probes we have considered up to now are distance related quantities. As an alternative route,
one can, in principle, use time\,-\,related quantities such as the lookback time \cite{CCFA04} or the age of old high redshift galaxies (OHRG) \cite{AL99,LA00,LJC09}. Unfortunately, at the moment, such a test can give only weak constraints since the data are too sparse and affected by large errors and possibly by systematic biases still to be investigated in detail. We have therefore not used this test in our analysis, but we check here a posteriori whether the HS and St $f(R)$ models are consistent with the age data. To this end, we follow \cite{LJC09} and consider the age of the Universe at redshift $z$\,:

\begin{equation}
t(z, {\bf p}) = t_H \int_{0}^{z}{\frac{dz'}{(1 + z') E(z', {\bf p})}}
\label{eq: tzdef}
\end{equation}
with $t_H = 9.78 h^{-1} \ {\rm Gyr}$ the Hubble time. Given a galaxy at redshift $z_G$, we can get an estimate of $t(z_G)$ as the sum of the age $t_G$ of the galaxy (inferred by fitting its observed spectrum to stellar population synthesis models) and the incubation time $t_{inc}$. This latter gives us an estimate of the amount of time between the beginning of structure formation and the actual formation time of the galaxy. We then use the OHRG sample assembled in \cite{LJC09} from the literature then available and, for each galaxy, estimate the {\it incubation time}\,:

\begin{figure}
\centering
\subfigure{\includegraphics[width=6.5cm]{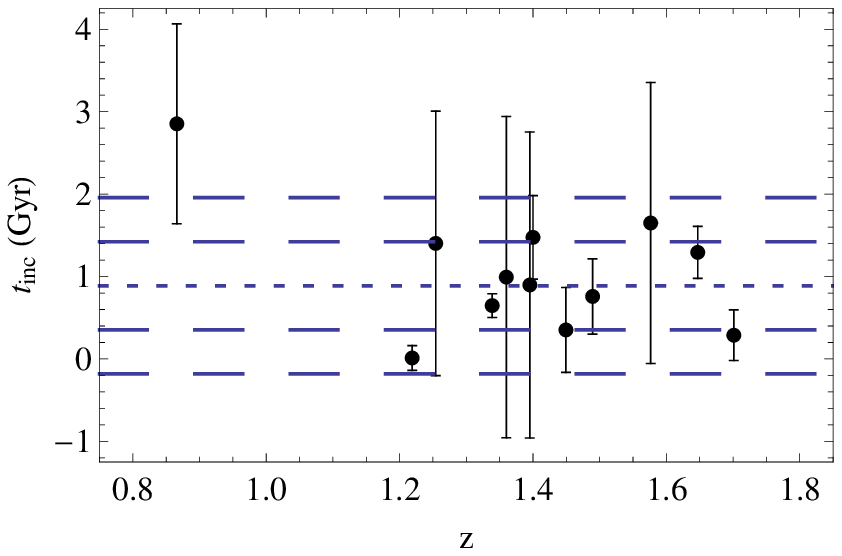}} \goodgap
\subfigure{\includegraphics[width=6.5cm]{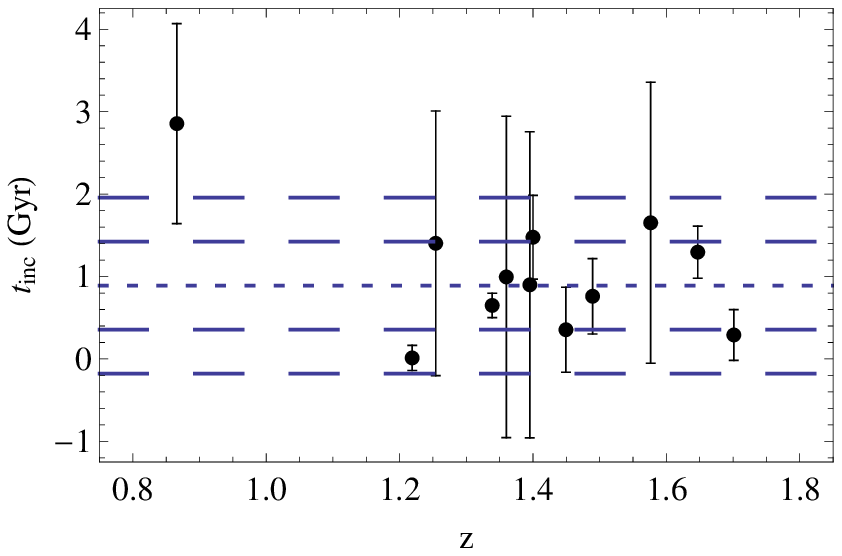}} \goodgap
\caption{Estimated incubation time for the HS (left) and St (right) $f(R)$ models for the galaxies in the sample given by \cite{LJC09}. Short and long dashed lines refer to the central value and $1$ and $2 \sigma$ ranges for $t_{inc}$ as estimated from stellar population synthesis models.}
\label{fig: inchsst}
\end{figure}

\begin{displaymath}
t_{inc} = t_{mod}(z_G) - t_{obs}(z_G)
\end{displaymath}
with the error naively estimated as\,:

\begin{displaymath}
\sigma_{inc} = \sqrt{\sigma_{mod}^2 + \sigma_{obs}^2} \ ,
\end{displaymath}
where the quantities with the underscript $mod$ are estimated evaluating $t(z_G)$ along the chains. Assuming that the formation redshift is approximately the same for all the galaxies in the sample (which is a reasonable approximation), the incubation time estimated from our model should be the same for all the objects in the sample. Moreover, its value should be in agreement with $t_{inc} = 0.8 \pm 0.4 \ {\rm Gyr}$ estimated based on stellar population evolutionary time \cite{tincest}. To this end, we therefore compute $t_{inc}$ from the chains obtained before fitting the HS and St models to the full dataset and plot the results in Fig.\,\ref{fig: inchsst}. As it is apparent, the errors are still quite large so that a definitive conclusion can not be safely drawn. Nevertheless, we note that, for all galaxies but that at the lowest redshift (which is however consistent within the $2 \sigma$ error bar), $t_{inc}$ is in agreement with the estimated value for both the HS and St models. Thus, we do not find any age problem for these $f(R)$ theories.

The age data are difficult to extend to higher redshift because of the prohibitively long exposure time needed to get a sufficiently high quality spectra of galaxies at $z > 2$. The situation is, however, better for quasars as it is well illustrated by the case of APM\,08729\,+\,5225. With a redshift $z = 3.91$, this object has an estimated age of $2$\,-\,$3 \ {\rm Gyr}$ \cite{HSK02} with a best fit age $t_{APM} = 2.1 \ {\rm Gyr}$ and a lower $1 \sigma$ limit $t_{APM} \ge 1.7 \ {\rm Gyr}$ \cite{F05}. Note that the incubation time is here not well defined, but it is likely to be very small. Asking that the age of the Universe at $z = 3.91$ is larger than the age of APM\,08729\,+\,5225 allows to set constraints on dark energy models \cite{apmtest1,apmtest2,apmtest3,apmtest4} and shows that, for many models (including the concordance $\Lambda$CDM), the condition $t(z = 3.91) = t_{3.91} > t_{APM}$ is not easy to fulfill. From our chains, we get\,:

\begin{displaymath}
t_{3.91} = 1.568_{-0.037 \ -0.075}^{+0.036 \ +0.078} \ \ {\rm Gyr} \ \ , \ \
t_{3.91} = 1.569_{-0.038 \ -0.070}^{+0.037 \ +0.077} \ \ {\rm Gyr} \ \  \ \ , \ \
\end{displaymath}
for the HS and St models, respectively. For both cases, the $95\%$ upper limit on $t_{3.91}$ is smaller than the $1 \sigma$ lower limit of $t_{APM}$ so that we get the same age problem as for the $\Lambda$CDM scenario. Although one can repeat our likelihood analysis explicitly including a prior on $t_{3.91}$, the agreement with the extensive dataset considered above makes us confident that the problem could reside in the uncertainties in dating high $z$ quasars. This can lead to bias high the estimate of $t_{APM}$ rather than in a failure of the $f(R)$ theories. As a further evidence in favor of this interpretation, one may note that it is actually quite difficult to get $t_{3.91} > t_{APM}$ for almost all the models in the literature.

We then consider the present day age of the Universe, $t_0 = t(z = 0)$, obtaining (in Gyr)\,:

\begin{displaymath}
t_{0} = 13.37_{-0.25 \ -0.54}^{+0.26 \ +0.56} \ \ {\rm Gyr} \ \  \ \ , \ \
t_{0} = 13.34_{-0.25 \ -0.46}^{+0.28 \ +0.55} \ \ {\rm Gyr} \ \  \ \ , \ \
\end{displaymath}
for the HS and St cases. Both these constraints agree well with previous ones in the literature (see, e.g., \cite{WMAP7} and refs. therein), but are smaller than the estimated age of old globular clusters in the Milky Way, $t_{GC} \simeq 14 \ {\rm Gyr}$ \cite{Pont98}. This discrepancy is very small % and can be likely solved adjusting the globular cluster age estimate. Indeed,
and, in fact, a different estimate of the globular cluster age, $t_{GC} = 12.6^{+3.4}_{-2.6} \ {\rm Gyr}$ \cite{KC03}, removes the discrepancy for both $f(R)$ models.

As a final general remark on these time based tests, we note that the constraints on both $t_{3.91}$ and $t_0$ turn out to be essentially the same for both $f(R)$ models considered. This is a further consequence of how very similar both models actually are. Indeed, for $z \ge 2$, both $f(R)$ functions are actually well approximated by $R - 2 \Lambda_{eff}$, while the constraints on each model parameters simply translate into the same effective $\Lambda$ term. As a consequence, discriminating among them is quite difficult both with distance and time related quantities because the smoothing of the evolution rate due to the integration further cancels out the subtle differences in the two $f(R)$ functions.

\section{The growth of perturbations}

As extensively shown here, both the HS and St models provide an evolutionary history that can be hardly distinguished from $\Lambda$CDM. This is not surprising given that $f(R)$ reduces to $f_{\Lambda}(R)$ for $R >> R_s$ with $R_s$ a characteristic curvature value depending on the model. Since the likelihood analysis points towards model with small values of $R_s$, the condition $R >> R_s$ is soon fulfilled for most of the redshift range probed (from $z = 0$ to the last scattering surface $z_{LS}$). In order to discriminate between the HS and St models (and, more generally, between $f(R)$ theories and dark energy), one must resort to the observables probing the growth of perturbations such as cosmic shear \cite{frWL,matteobis,stefano,stefanobis}. As a preliminary analysis, we will consider here some theoretical constraints on quantities related to the growth of structures in the two $f(R)$ models we are considering.

\subsection{The growth index}

We start by considering here the growth index of perturbations. In the subhorizon limit, a primordial matter perturbation $\delta = \delta \rho_M/\rho_M$ increases according to \cite{T07}\,:

\begin{equation}
\ddot{\delta} + 2 H \dot{\delta} - 4 \pi {\cal{G}}_{eff}(a, k) \rho_M \delta = 0
\label{eq: growtheq}
\end{equation}
with $k$ the wavenumber and

\begin{equation}
{\cal{G}}_{eff}(a, k) = \frac{G}{f^{\prime}(R)} \frac{1 + 4 (k^2/a^2) [f^{\prime \prime}(R)/f^{\prime}(R)]}{1 + 3 (k^2/a^2) [f^{\prime \prime}(R)/f^{\prime}(R)]}
\label{eq: geffdef}
\end{equation}
the scale dependent effective gravitational constant. Note that, differently from standard GR, the growth equation now depends on the scale $k$ so that we will have a scale\,-\,dependent growth of perturbations. Rather than solving Eq.(\ref{eq: growtheq}), we can solve for the growth rate $g = d\ln{\delta}/d\ln{a}$. Using the redshift $z$ as variable, Eq.(\ref{eq: growtheq}) leads to\,:

\begin{equation}
\frac{dg(z)}{dz} + \left [ \frac{1}{2} \frac{d\ln{E^2(z)}}{d\ln{(1 + z)}} - (2 + g) \right ] \frac{g(z)}{1 + z}
+ \frac{3}{2} \frac{\Omega_M (1 + z)^2}{E^2(z)} \frac{{\cal{G}}_{eff}(k, z)}{G} = 0
\label{eq: gzeq}
\end{equation}
with the initial condition $g(z \simeq 100) = 1$, i.e., we ask that at very high $z$ the growth rate is the same as the one for a matter only universe in GR. If GR holds, a useful parametrization for the growth rate is given by \cite{gi1,gi2,gi3}\,:

\begin{equation}
g(z) = \left [ \frac{\Omega_M (1 + z)^3}{E^2(z)} \right ]^{\gamma}
\label{eq: defgamma}
\end{equation}
where $\gamma$ is referred to as the growth index. For dark energy models with constant EoS, it is $\gamma = 3(w_0 - 1)/(6 w_0 - 5)$ which give $\gamma = 6/11 \simeq 0.545$ for $\Lambda$CDM.

For $f(R)$ theories, Eq.(\ref{eq: gzeq}) is scale dependent so that we must check whether the approximating formula (\ref{eq: defgamma}) holds at each wavenumber $k$. As a first result, we have fitted the approximated relation\,:

\begin{figure*}
\centering
\subfigure{\includegraphics[width=6.5cm]{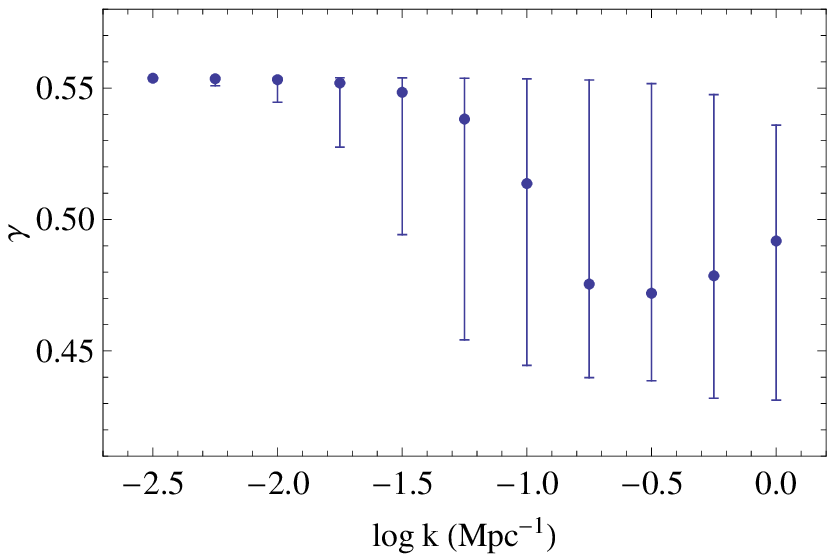}} \goodgap
\subfigure{\includegraphics[width=6.5cm]{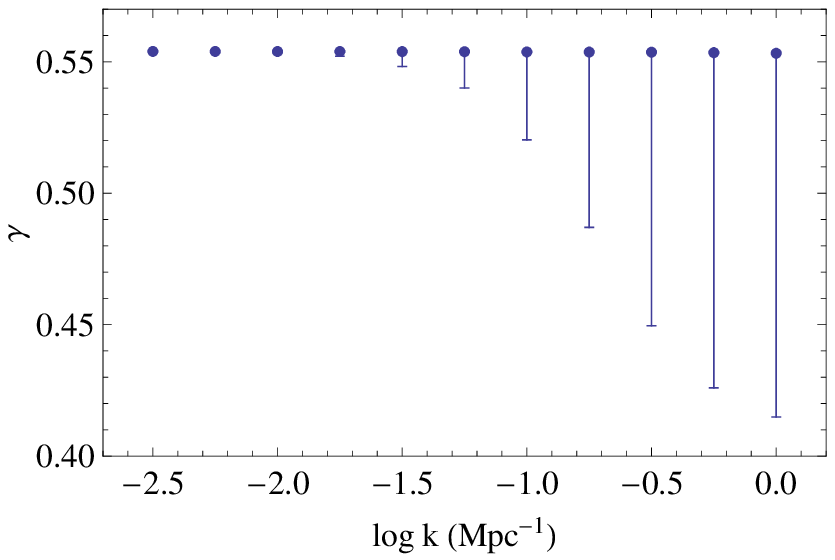}} \goodgap \\
\caption{Growth index $\gamma$ as function of $\log{k}$ for the HS (left) and St (right) models. Here and in the following plots, the bars refer to the $68\%$ CL with the point denoting the median value.}
\label{fig: giplots}
\end{figure*}

\begin{equation}
\log{g(z, k)} = \log{g_0} + \gamma \log{\left [ \frac{\Omega_M (1 + z)^3}{E^2(z)} \right ]}
\label{eq: gammafr}
\end{equation}
to the solution of Eq.(\ref{eq: gzeq}) for both the HS and St models and different values of the model parameters and wavenumber. We indeed find that Eq.(\ref{eq: gammafr}) fits the numerical solution with a rms error which, although increasing with $k$ (up to $k \simeq 0.3 \ {\rm Mpc}^{-1}$), is always smaller than $2\%$ ($1\%$) for the HS (St) model. Moreover, $g_0$ is always unity within less than $0.1\%$ consistent with the prediction of Eq.(\ref{eq: defgamma}). That such an approximation for the growth factor $g(z, k)$ indeed works is actually not a surprise considering the results in \cite{Gan09,Tsu09,Moto10} which have already investigated this issue for the St\,-\,like model. We nevertheless note that their approach is slightly different since, in \cite{Gan09}, the authors integrate an equation for $g(z, k)$ as a function of $\Omega_M(z)$, while an equation for $\gamma(z)$ is derived and solved in \cite{Tsu09,Moto10}. On the contrary, we here define $\gamma(z, k)$ as a fitting parameter to a numerically derived $g(z)$ vs $\Omega_M(z)$ relation for any given $k$. As a consequence, the different results can not be quantitatively compared.

It is worth stressing that $\gamma$ depends on $k$ so that it is interesting to see whether this signature may be used to discriminate between the two classes of $f(R)$ theories and between them and dark energy. To take into account errors on the model parameters, we have therefore fitted (\ref{eq: gammafr}) to the numerical solution of (\ref{eq: gzeq}) for a given $k$ and for all the points in the chain. The results are shown in Fig.\,\ref{fig: giplots} for both the HS and St models. Here (and in the following plots), we report the $68\%$ confidence range obtained by evaluating $\gamma(\log{k})$ along the final chain with the point denoting the median value. Note that the resulting error bars are typically asymmetric as a consequence of the non Gaussian distribution of the $\gamma$ values. Moreover, for the St model, the $\gamma$ distribution is actually quite narrow, but some outliers generate long tails towards small $\gamma$ thus leading to strongly asymmetric error bars.

There are two remarkable lessons which can be drawn from these plots. First, $\gamma$ is well determined for $\log{k} \le -1.5$ for both $f(R)$ theories, while a stronger sensitivity to the model parameters takes place for larger $k$ values. Second, the scale dependence of $\gamma$ is negligible at all for $\log{k} \le -1.5$ with both models having $\gamma = 0.55$ within an excellent accuracy. On the contrary, on smaller scales (i.e., larger $k$), the St model has still a growth index essentially constant, while a significative scale dependence is present for the HS model with $\gamma$ becoming smaller than the $\Lambda$CDM value. This characteristic features suggest that a possible way to discriminate between the two $f(R)$ models is to devise a method able to measure $g(z)$ at different scales and redshifts which could, however, be a formidable observational task. On the other hand, on the theoretical side, we should correct our estimates of $\gamma$ for $\log{k} > -1.5$ to take into account deviations from the linear regime\footnote{One can roughly consider the linear regime valid up to scale where the variance of the matter power spectrum is of order unity. For a $\Lambda$CDM model, such an upper limits turns out to be $k \sim 0.15 h \ {\rm Mpc}^{-1}$, i.e. $\log{k} \sim -1.0$ for the $h \sim 0.7$ as we find here. In $f(R)$ gravity, this limit can be also smaller depending on which recipe is used to correct the linear power spectrum for the nonlinear effects (which can boost the growth of perturbations by several percents) so that, as a conservative choice, we have warn the reader to take with caution the results for $\log{k} > -1.5$ and exclude them from successive fits.} which can take place at such large $k$.

\begin{figure*}
\centering
\subfigure{\includegraphics[width=4.5cm]{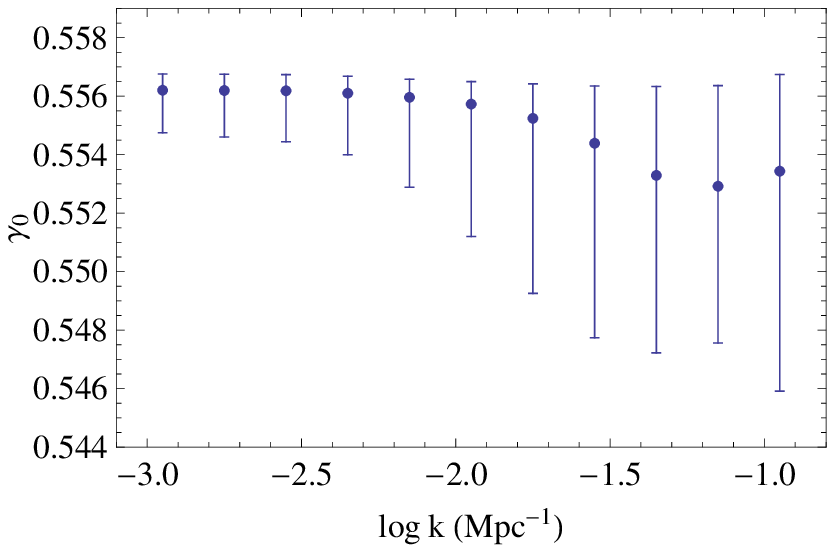}} \goodgap
\subfigure{\includegraphics[width=4.5cm]{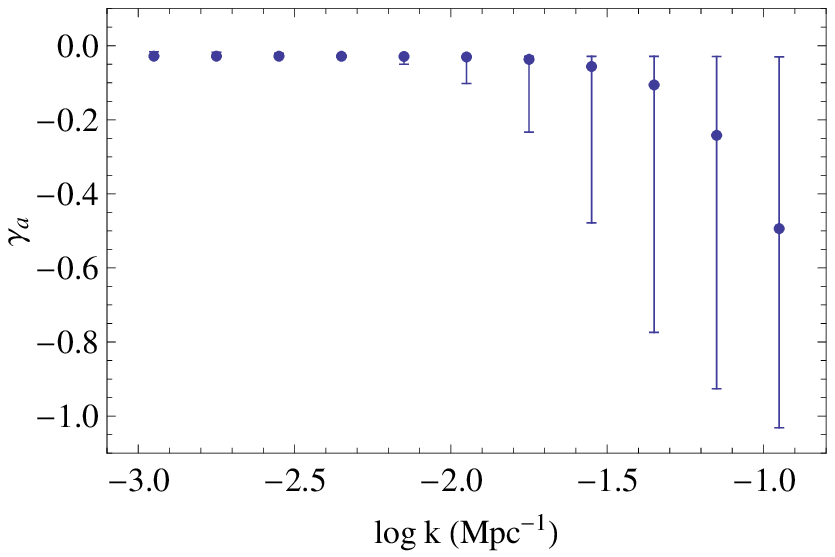}} \goodgap
\subfigure{\includegraphics[width=4.5cm]{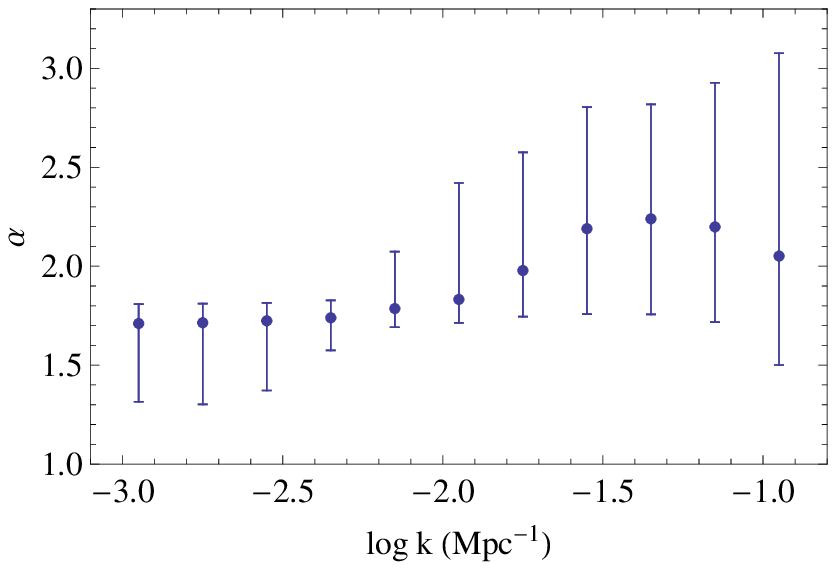}} \goodgap \\
\subfigure{\includegraphics[width=4.5cm]{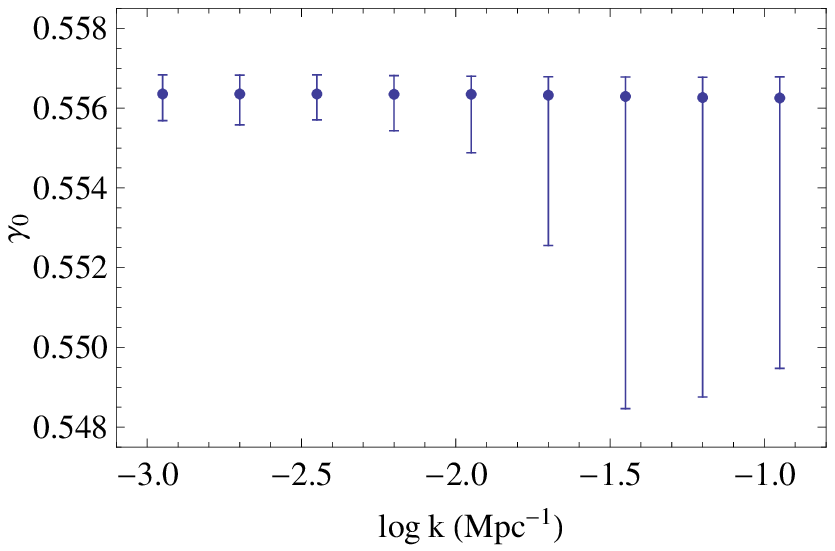}} \goodgap
\subfigure{\includegraphics[width=4.5cm]{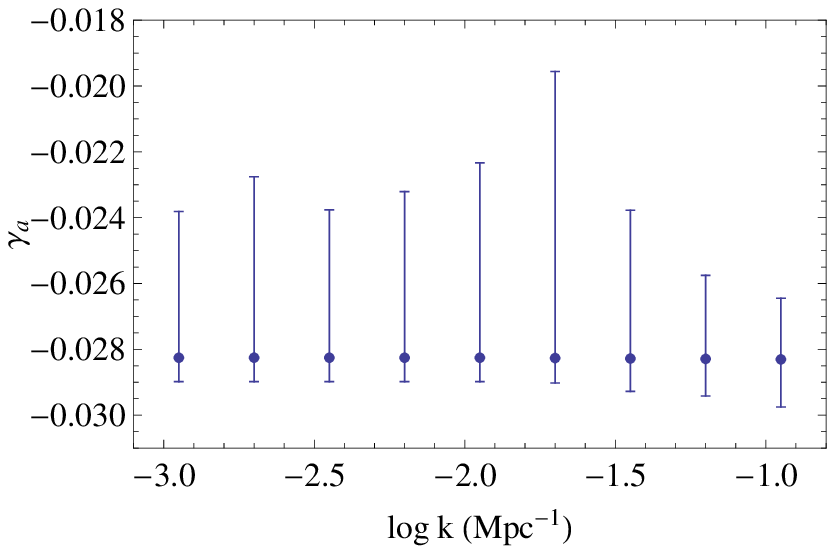}} \goodgap
\subfigure{\includegraphics[width=4.5cm]{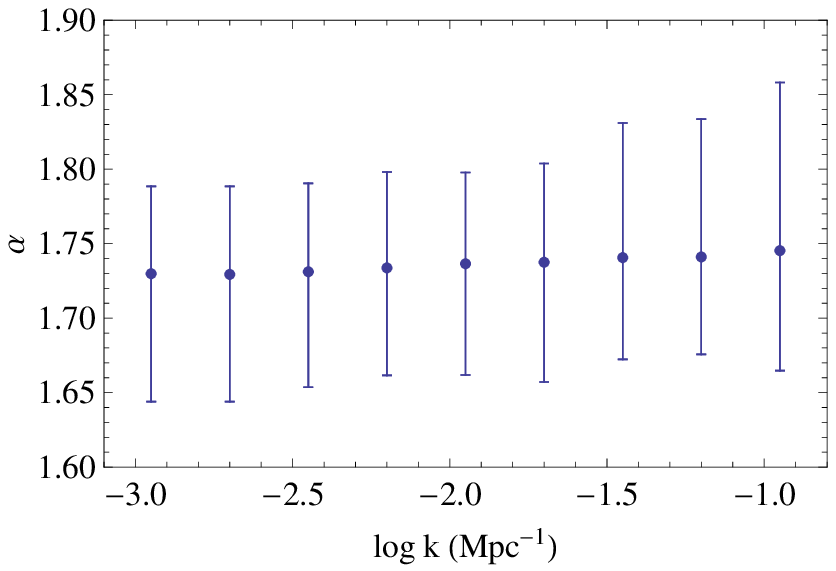}} \goodgap \\
\caption{Growth index fitting parameters vs $\log{k}$ for the HS (top) and St (bottom) models.}
\label{fig: gbfits}
\end{figure*}

An alternative parametrization for the growth index has been recently proposed in \cite{GB11} to allow for a redshift dependent growth index. Indeed, one still relies on Eq.(\ref{eq: defgamma}), but now $\gamma$ is assumed to scale with $z$ as $\gamma(z) = \gamma_0 + \gamma_a z/(1 + z)$. In order to allow for a larger flexibility, we generalize the \cite{GB11} proposal to $\gamma(z) = \gamma_0 + \gamma_a z (1 + z)^{-\alpha}$ and fit the parameters $(\gamma_0, \gamma_a, \alpha)$ to the $g(z)$ vs $\Omega_M(z)$ relation for each of the point along the chains for both the HS and St models. We then plot the median value and the $68\%$ confidence ranges in Fig.\,\ref{fig: gbfits} as function of $k$ considering only scales $\log{k}\le -1.5$ in order to avoid nonlinear effects. As a first general comment, we find that the approximation works quite well: it reproduces the input $g(z)$ with an error smaller than $1\%$ for all the set of parameters. Moreover, the value of $\alpha$ is larger than unity and indicates that the redshift evolution is actually more rapid than what is implemented in the \cite{GB11} parametrization. Although the error bars are significantly smaller for the St than for the HS model, the median values of both $\gamma_0$ and $\gamma_a$ are very similar in the two models up to $\log{k} \sim -1.5$, while more negative values are obtained for the HS model on smaller scales. It is worth stressing that, although small, $\gamma_a$ takes a non null value and, in particular, $|\gamma_a| > 0.02$ in agreement with \cite{PG08} who, based on an analysis of the growth of perturbations in GR dark energy models, have argued that $|\gamma_a| > 0.02$ could be a signature for modified gravity. Finally, whereas $\alpha \simeq 1.735$ independent of the scale is a good approximation for the St model, this is not the case for the HS $f(R)$ theory that shows a clear increasing trend with $k$. Note that, since $\gamma_a \sim -0.028$ in both cases, this result show that the growth index decreases with $z$ in a faster way for the HS than the St model. Although this could suggest a possible way to discriminate between the two models, observationally detecting the redshift and scale dependence of the growth index is a rather unrealistic task.

\subsection{The growth factor}

The growth index $\gamma$ is not directly observable, but it must be inferred from measurements of both the matter density parameter at a given redshift and the growth factor $g(z)$ at the same $z$. This latter quantity may be estimated from redshift space distortions in the galaxy power spectra at different $z$ \cite{K87,H98}. Actually, what is directly measured is the ratio $g(z)/b(z)$ with $b(z)$ the linear bias needed to convert the observed galaxy power spectrum to the matter one. If, as a first rough approximation, we assume that the bias is the same for both $f(R)$ and $\Lambda$CDM models, we can then solve Eq.(\ref{eq: gzeq}) and compare the predicted growth factor with the observed one.

\begin{figure*}
\centering
\subfigure{\includegraphics[width=6.5cm]{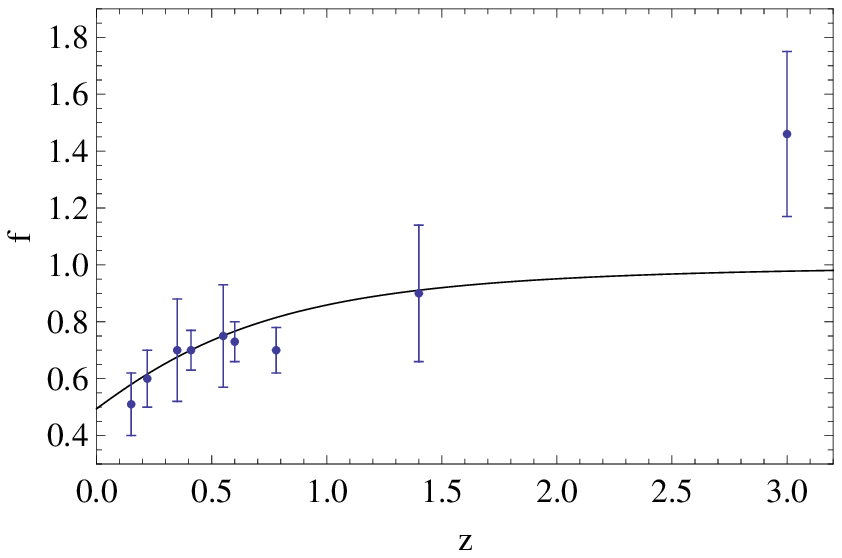}} \goodgap
\subfigure{\includegraphics[width=6.5cm]{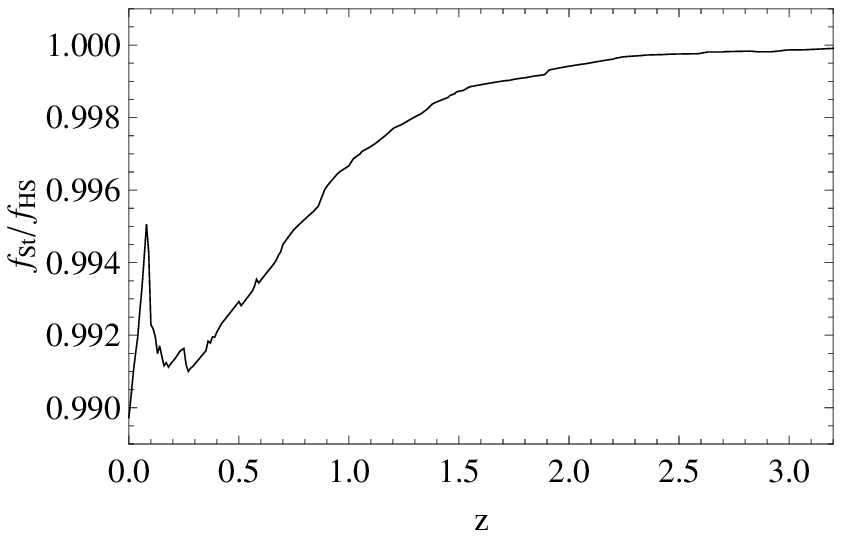}} \goodgap \\
\caption{Growth factor as function of $z$. {\it Left.} $g(z)$ vs $z$ for the HS model with $k = 0.01 \ {\rm Mpc^{-1}}$ with superimposed the observed data. {\it Right.} Ratio between the growth factor for the St and HS models evaluated at $k = 0.01 \ {\rm Mpc^{-1}}$. In both panels, we plot $g(z)$ as inferred from the median of the values along the corresponding chains. Note that the spike in the right panel at low $z$ is only a numerical artifact and not a physical feature.}
\label{fig: fzplots}
\end{figure*}

In order to take care of the uncertainties on the model parameters, we solve Eq.(\ref{eq: gzeq}) for all the points in the chain and finally plot the median result having checked that the $95\%$ confidence range for $g(z)$ at any given $z$ is negligibly small compared to the statistical uncertainties. The left panel in Fig.\,\ref{fig: fzplots} shows the result for the HS model setting $k = 0.01 \ {\rm Mpc^{-1}}$, but the plot is the same as long as we remain in the linear regime (roughly, $k \le 0.1 \ {\rm Mpc^{-1}}$). We overplot the $g(z)$ data as reported in \cite{NP08} assembling the different measurements then available and here complemented by the recent measurements from the WiggleZ survey \cite{B11}. As it is apparent, there is a very good match between the model and the data up to $z = 1.5$, while the point at $z = 3$ (derived from Ly$\alpha$ forest power spectrum) appears to disagree. However, this latter measurement might be biased by some error in the $b(z)$ estimate at such large redshift; in fact, even $\Lambda$CDM, which is successful in many other resepcts, is unable to fit this point. Although such an agreement with the growth factor data is surely encouraging, it is worth stressing that they are based on the assumption of same linear bias for both GR based $\Lambda$CDM model and modified gravity $f(R)$ scenarios. Therefore, any conclusion drawn from Fig.\,\ref{fig: fzplots} must be taken {\it cum grano salis}.

Finally, we note that the growth factor for the St model also is consistent with the $g(z)$ data given the nearly coincidence with the HS model prediction. Indeed, as the right panel in Fig.\,\ref{fig: fzplots} show, the two growth factors closely track each other along the full redshift range with a maximum deviation of $\sim 1\%$ at very low $z$ which is hard to observationally detect with both present and future data. We therefore argue that the growth factor alone can not be used to discriminate between the two classes of $f(R)$ theories we are here investigating.

\subsection{The deviations from the GR growth of perturbations}

\begin{figure*}
\centering
\subfigure{\includegraphics[width=6.5cm]{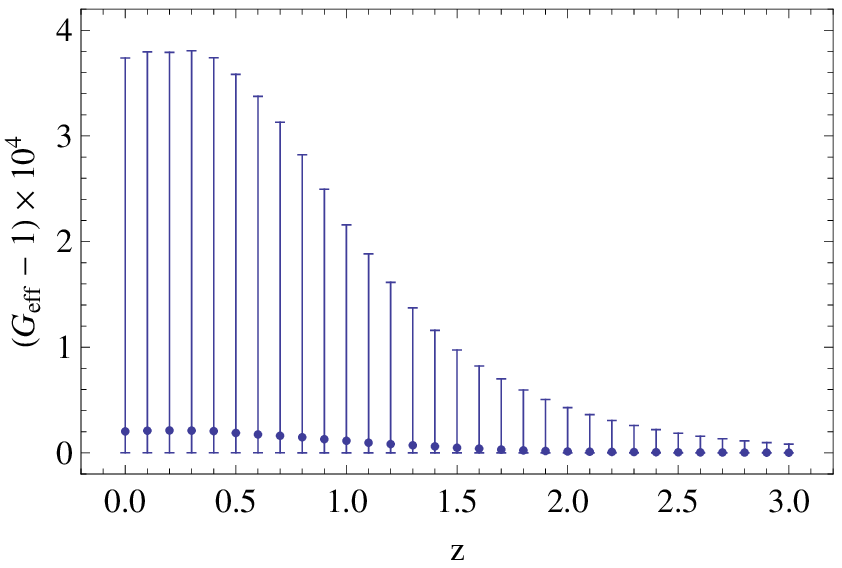}} \goodgap
\subfigure{\includegraphics[width=6.5cm]{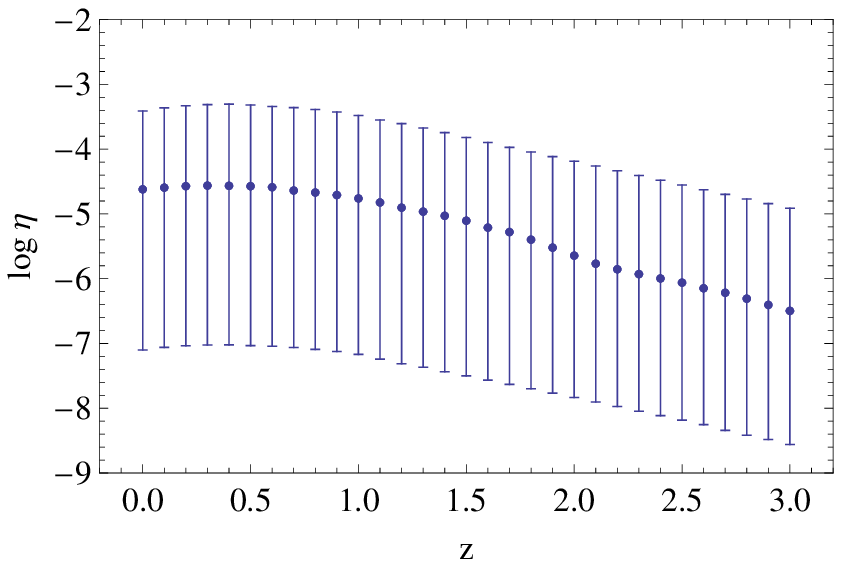}} \goodgap \\
\subfigure{\includegraphics[width=6.5cm]{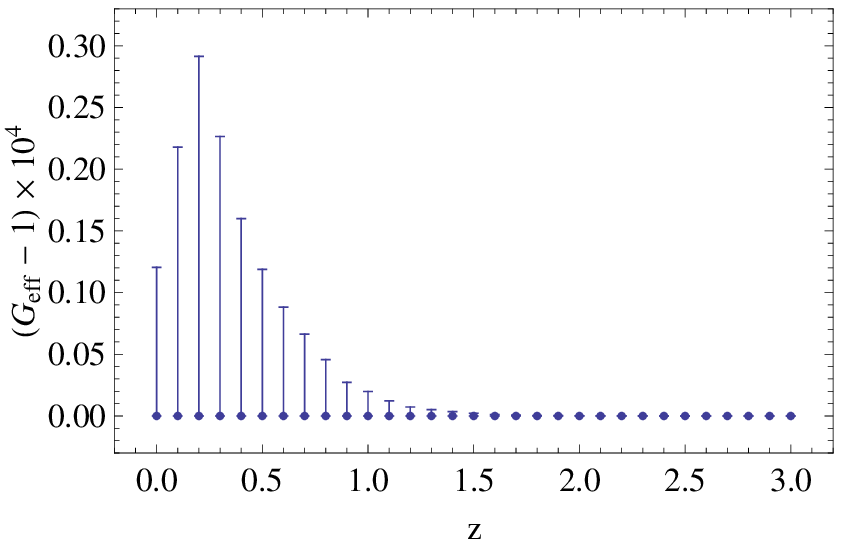}} \goodgap
\subfigure{\includegraphics[width=6.5cm]{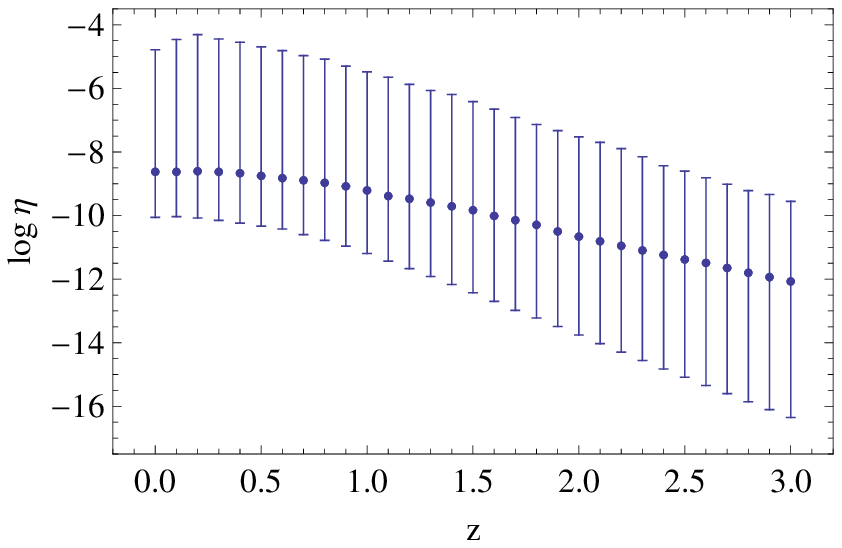}} \goodgap \\
\caption{Deviations from the GR growth of perturbations in terms of ${\cal{G}}_{eff}/G_N$ (left) and $\log{\eta}$ (right) for the HS (top) and St (bottom) models setting $k = 0.001 \ {\rm Mpc^{-1}}$.}
\label{fig: geffplots}
\end{figure*}

So far, we have only considered the growth of structures as probed by observable quantities related to the matter power spectrum. Actually, the growth of structures can also be probed from the point of view of the metric potentials entering the perturbed line element\,:

\begin{displaymath}
g_{\mu \nu} dx^{\mu} dx^{\nu} = -(1 + 2 \Phi) dt^2 + (1 + 2 \Psi) d{\bf x}^2 \ .
\end{displaymath}
While $\Phi$ enters the Poisson equation, which in Fourier space reads\,:

\begin{displaymath}
\Phi_k(a) = - 4 \pi {\cal{G}}_{eff}(k, a) (k^2/a^2) \rho_m \delta_k(a) \ ,
\end{displaymath}
the other metric potential $\Psi$ enters the weak lensing potential $\Upsilon = -(\Phi - \Psi)/2$ and is related to $\Phi$ through the parameter $\eta = -(\Phi + \Psi)/\Phi$. In GR, one has ${\cal{G}}_{eff}/G_N = 1$ and $\eta = 0$ (i.e., no anisotropic stress), while, for $f(R)$ theories, ${\cal{G}}_{eff}(k, a)$ is given by Eq.(\ref{eq: geffdef}) and

\begin{equation}
\eta(k, a) = \frac{2 (k^2/a^2) [f^{\prime \prime}(R)/f^{\prime}(R)]}{1 + 2 (k^2/a^2) [f^{\prime \prime}(R)/f^{\prime}(R)]} \ .
\label{eq: defeta}
\end{equation}
It is then interesting to estimate the theoretical predictions for ${\cal{G}}_{eff}(k, a)$ and $\eta(k, a)$ in order to see whether they can help to discriminate between the proposed modified gravity scenarios and the concordance $\Lambda$CDM. ${\cal{G}}_{eff}$ and $\eta$ are shown in Fig.\,\ref{fig: geffplots} where the central value refers to the median and the plotted bar denote $68\%$ confidence range as inferred from the set of parameters along the chain. Note that we have set $k = 0.001 \ {\rm Mpc^{-1}}$, but the results can be easily scaled to other values. We also stress that the strong asymmetry of the confidence range around the median is a consequence of how ${\cal{G}}_{eff}$ is defined. Indeed, in the high\,-\,$z$ regime, both $f^{\prime}(R)$ and $f^{\prime \prime}(R)$ become negligibly small and one recovers ${\cal{G}}_{eff}/G_N = 1$, while, in the low\,-\,$z$ limit, all the relevant terms are positive thus leading to ${\cal{G}}_{eff}/G_N > 1$. As a final result, ${\cal{G}}_{eff}/G_N$ can not take values smaller than 1 along the chains thus motivating the strongly asymmetric confidence ranges.

The median values for ${\cal{G}}_{eff}/G$ are close to 1 as expected. Indeed, over most of the parameter space allowed by the data, both the HS and St Lagrangians are quite similar to GR\,+\,$\Lambda$ so that we indeed expect to recover an effective gravitational constant equal to the Newtonian $G$. Moreover, since the set of parameters have to combine in such a way to reproduce the data as well as the $\Lambda$CDM model, the inferred confidence ranges for ${\cal{G}}_{eff}/G$ are quite narrow and deviations from unity become more and more unlikely as $z$ increases since, in this regime, both $f(R)$ models reduce to an effective $\Lambda$CDM. A similar qualitative discussion also explains the behavior of $\eta(z)$ so that deviations of the lensing potential from GR are smaller and smaller as $z$ increases. We finally note that, for the St model, ${\cal{G}}_{eff}(k, a)$ and $\eta(k, a)$ are always closer to the GR values than for the HS model. We can therefore anticipate that the cosmic shear power spectrum is unable to discriminate between the St and $\Lambda$CDM models. This is, indeed, what we find in \cite{stefano} when tomography is not used; on the contrary the HS model can give rise to a detectable signature.

\section{Conclusions}

As soon as the observational and conceptual problems related to the cosmological constant and other dark energy scenarios became pressing, modification of the gravity sector of Einstein field equations immediately appeared as an interesting alternative explanation of the observed cosmic accelerated expansion. Fourth\,-\,order gravity theories then stand out as the most immediate generalization of Einstein GR since they just encoded all the deviations into a single analytic function $f(R)$. As the problem of acceleration was solved in this framework, a new problem came out, namely how to choose a functional expression for $f(R)$ which is able to speed up the expansion and, at the same time, does not violate the local tests of gravity and turns off its effect in the early Universe where GR appears to work correctly.

From a mathematical point of view, one has to look for an $f(R)$ expression satisfying some constraints imposed to both fulfill the Solar System tests and recover GR during the nucleosynthesis era which is indeed what the HS and St models efficiently do, postulating that $f(R)$ is given by Eqs.(\ref{eq: frhs}) and (\ref{eq: frst}), respectively. Our aim here was then to test whether these two well behaved models actually fit the data which suggest the accelerated expansion of the Universe. To this end, we have considered the background evolution as traced by the SNeIa and GRBs Hubble diagrams, the $H(z)$ data from the differential age method applied to passively evolving red galaxies, the BAO constraints on $r_s(z_d)/D_V(z)$ and the WMAP7 distance priors. The combined dataset allows us to probe both the late Universe ($z < 1.5$), the matter dominated era (through GRBs) and the last scattering surface epoch. We indeed find that both the HS and St models are able to fit this extended dataset with values of $(\Omega_M, h, q_0, j_0)$ in good agreement with previous results in the literature. We can therefore conclude that both the HS and St models are not only theoretically viable $f(R)$ proposals, but also observationally motivated choices for a fourth\,-\,order gravity Lagrangian.

An unpleasant outcome of our analysis is represented by the very weak constraints we give on the $f(R)$ parameters and the impossibility to discriminate between the two models. This is a consequence of the two $f(R)$ functional expressions having been tailored to reduce to the $\Lambda$CDM Lagrangian in the high curvature regime. Therefore, the background evolution is almost the same and only weak constraints can be set even using extraordinary precise data if these only probe the background evolution. Major improvements in both narrowing down the confidence ranges and in the possibility to discriminate not only between the HS and St models, but, more generally, between dark energy and $f(R)$ theories, are expected from the analysis of the growth of perturbations. Indeed, even if one can tailor the $f(R)$ parameters in order to closely mimic the same expansion history of a given dark energy model, the evolution of the density perturbations can be rather different. In particular, this has a strong impact on both the power spectrum and halo statistics \cite{frstab,PS08,OLH08,SLOH09,BJ09} and the weak lensing signals \cite{frWL,matteobis,stefano,stefanobis} and offers the possibility to compare predictions with data and to severely constrain the viability of $f(R)$ theories.

As a preliminary investigation, we have here derived the growth index $\gamma$ for the HS and St $f(R)$ theories showing that the usual parametrization of the growth rate still holds provided a scale dependent $\gamma$ is used. However, such a scale dependence is quite weak in the deep linear regime (say, e.g., $k < 0.01$), while the stronger variation for larger $k$ has to be confirmed in the nonlinear regime. Moreover, we have also improved the usual growth index parametrization allowing for a redshift dependence which turns out to be stronger for the HS than for the St model. Unfortunately, using this signature to discriminate between the two classes of $f(R)$ theories is theoretically possible, but practically quite difficult (if not impossible) to implement. As an alternative way, one can directly resort to the growth factor data as estimated from the redshift distortions of the galaxy power spectrum. Assuming that the linear bias is the same as in the GR framework, we have shown that both $f(R)$ models agree quite well with the present day data so that it is not possible to discriminate among them and GR. Actually, one should first check that the bias is indeed the same in the two different frameworks considering that, in $f(R)$ theories, not only the growth of structure is different, but also the gravitational potential which plays a key role in the formation of galaxies and hence in the determination of their bias function.

All these preliminary investigations may be considered as a first step towards an improved analysis of these two classes of $f(R)$ theories. It is indeed the combination of extended background evolution tracers (such as SNeIa and GRBs Hubble diagram, BAO and CMB distance priors) and structure growth probes (including galaxy power spectrum and cosmic shear) that will finally tell us whether the observed cosmic speed up has been the first
evidence of a new fluid, as mysterious as fascinating, or of new physics in the gravity sector, as unexpected as challenging.

\acknowledgments

VFC warmly thanks S. Capozziello and M.G. Dainotti for working together on GRBs. VFC is supported by Agenzia Spaziale Italiana (ASI). SC and AD gratefully acknowledge partial support from INFN grant PD51 and the PRIN-MIUR-2008 grant \verb"2008NR3EBK_003" ``Matter-antimatter asymmetry, dark matter and dark energy in the LHC era''. This research has made use of NASA's Astrophysics Data System.

\end{document}